\documentclass[fleqn,usenatbib]{mnras}

\usepackage{newtxtext,newtxmath}

\usepackage[T1]{fontenc}

\DeclareRobustCommand{\VAN}[3]{#2}
\let\VANthebibliography\thebibliography
\def\thebibliography{\DeclareRobustCommand{\VAN}[3]{##3}\VANthebibliography}

\usepackage{graphicx}	
\usepackage{amsmath}	
\usepackage{pdflscape}
\usepackage{rotating}

\newcommand{\XY}[2]{\left[\textrm{#1/#2}\right]}
\newcommand{\FeH}{\XY{Fe}{H}}

\newcommand{\kms}{km\,s$^{-1}$}
\newcommand{\Teff}{T_\textrm{eff}}
\newcommand{\logg}{\log g}

\newcommand{\vmic}{v_\textrm{mic}}
\newcommand{\Msol}{\text M_\odot}

\title[Spectroscopic analysis of NGC 1846]{A high-resolution spectroscopic search for multiple populations in the 2 Gyr old cluster NGC 1846}

\author[W. S. Oh et al.]{
W. S. Oh,$^{1}$$^{,2}$\thanks{E-mail: weishen.oh@anu.edu.au}
T. Nordlander,$^{1}$$^{,2}$
G. S. Da Costa$^{1}$$^{,2}$
and A. D. Mackey$^{1}$$^{,2}$
\\
$^{1}$Research School of Astronomy and Astrophysics, Australian National University, Canberra, ACT 2611\\
$^{2}$ARC Centre of Excellence for All Sky Astrophysics in 3 Dimensions (ASTRO 3D), Australia
}

\date{Accepted XXX. Received YYY; in original form ZZZ}

\pubyear{2022}

\begin{document}
\label{firstpage}
\pagerange{\pageref{firstpage}--\pageref{lastpage}}
\maketitle

\begin{abstract}
We present detailed C, O, Na, Mg, Si, Ca, Ti, V, Fe, Zr, Ba, and Eu abundance measurements for 20 red giant branch (RGB) stars in the LMC star cluster NGC 1846 ([Fe/H] = -0.59). This cluster is 1.95 Gyr old and lies just below the supposed lower age limit (2 Gyr) for the presence of multiple populations in massive star clusters. Our measurements are based on high and low-resolution VLT/FLAMES spectra combined with photometric data from HST. Corrections for non-local thermodynamic equilibrium effects are also included for O, Na, Mg, Si, Ca, Fe and Ba. Our results show that there is no evidence for multiple populations in this cluster based on the lack of any intrinsic star-to-star spread in the abundances of Na  and O: we place 95\,\% confidence limits on the intrinsic dispersion for these elements of $\le 0.07$ and $\le 0.09$\,dex, respectively. However, we do detect a significant spread in the carbon abundances, indicating varying evolutionary mixing occurring on the RGB that increases with luminosity. Overall, the general abundance patterns for NGC 1846 are similar to those seen in previous studies of intermediate-age LMC star clusters and field stars.                    
\end{abstract}




\section{Introduction}

It has been known for decades that almost every globular cluster (GC) in the Milky Way possesses multiple stellar populations. This refers to a cluster having two main stellar groups: a first generation (1G),
consisting of stars that are chemically similar to halo stars, and a second generation (2G), consisting of stars rich in He, N, Na and Al, and poor in C, O and Mg with respect to 1G. (\citealt{Gratton2012}; \citealt{Piotto2015}). 
UV and optical photometric data allows these stellar groups to be differentiated using colour-magnitude diagrams (CMD). Indeed, past studies have shown distinct CMD features for a large number of clusters ($\sim 60$) such as multiple red-giant branches (RGBs), sub-giant branches (SGBs) and even main sequences (MSs) (\citealt{Milone2009}; \citealt{Milone2017}), indicating He and N variations \citep{Milone2018}. These findings have been complemented by spectroscopic observations indicating star-to-star variations in light elements. In particular, they occur in the form of abundance anti-correlations between Na and O, Mg and Al, and C and N \citep{Bastian2018}.

However, the origin and mechanism behind these variations are not well understood, since they are not predicted by the basic theory of star cluster formation. Various attempts have been made to explain the production of such abundance patterns, and a popular theory involves the processing of first-generation stellar material at high temperatures, with the processed material then incorporated into a second generation of star formation via a suitable gas reservoir, mixed with some amount of unprocessed material with 1G composition (e.g. \citealt{DErcole2008}; \citealt{Conroy2011}). Examples of such processing can be found in intermediate-mass asymptotic giant branch (AGB) stars and massive rotating stars. In both cases, enriched material is brought up to the stellar surface where it can be released into the intracluster medium. The mass lost from these stars would form the gas reservoir that is needed to form  a second generation of stars with the 2G light-element abundance patterns.\par

Unfortunately, these and other proposed theories to date have been shown to have at least one fundamental flaw (\citealt{Renzini2015}; \citealt{Bastian2018}). For example, the two key models (AGB enrichment and fast rotating massive star enrichment) mentioned above both require the formation of multiple generations of stars and hence star formation spanning some extended interval. However, the maximum internal age dispersion observed in young massive star clusters of $\sim$30 Myr \citep{DeMarchi2011} is not sufficient for intermediate-mass stars to evolve to the AGB and start polluting the next generation, as this typically takes around 40-160 Myr. On the other hand, this age spread is too large for the small time interval ($\sim$6 Myr) required between the pollution from massive rotating stars and their supernova explosions \citep{Gratton2012}. Since none of the proposed models have been able to reproduce the main observational properties of multiple stellar populations without making ad hoc assumptions, a self-consistent explanation of the physical processes responsible for the multiple populations phenomenon is lacking, as well as an understanding of which (if any) cluster properties control whether a GC will host chemical anomalies or not (\citealt{Martocchia2018}; \citealt{Milone2019}).  \par

Past studies have shown that the role of the estimated initial cluster mass is an important factor in determining whether clusters display multiple populations \citep{Milone2019}. Abundance inhomogeneities are rarely seen in clusters with present-day masses less than $\sim10^{5}$ solar masses \citep{Gratton2012}, which explains why most of the Milky Way clusters that display multiple populations are globular clusters.

Another theory that has been studied in the past is the connection between chemical anomalies and the presence of an extended main-sequence turn-off (eMSTO) in CMDs of star clusters. The eMSTO feature is observed in young and intermediate-age massive clusters (20 Myr - 2 Gyr) (\citealt{Mackey2008}; \citealt{Bastian2018}; \citealt{Milone2018a}), and was initially hypothesised to be due to internal age spreads of up to a few hundred Myr, as predicted by some of the previously-discussed models for the formation of light element abundance variations in globular clusters. While multiple populations in GCs have sometimes been inferred purely from the presence of broadened or split main-sequences and turn-offs in the CMDs of young/intermediate-age clusters, these features are not necessarily associated with chemical abundance variations. Recent work by \cite{Kamann2020} (but also see the discussion in \citealt{Bastian2018}) has shown that stellar rotation might be a significant factor in causing features such as the eMSTO to form. Rotation alters the internal stellar structure, because the centrifugal support and extra mixing in the core changes its hydrostatic equilibrium compared to that of a non-rotating star of the same mass and composition. These factors cause the evolutionary path of the rotating star in the CMD to vary in temperature and colour relative to the equivalent non-rotating star. Hence, the relationship with multiple populations in younger clusters (if any) remains unknown. \par

Therefore, to determine the leading factor for the presence of multiple populations in star clusters, and to better understand the timescale of the multiple population process, one key method is to look for multiple populations in younger populous clusters in nearby galaxies \citep{Gratton2019}, since there are no such clusters in the Milky Way \citep{Portegies2010}. An example of such objects are the massive intermediate-age (2-8 Gyr) Magellanic Cloud clusters. These are the closest examples of systems with globular cluster-like masses but much younger ages, providing direct snapshots of the cluster formation and evolution process at different times. \par

Past studies have shown that LMC clusters are also found with an observed splitting or spread in the subgiant and red giant branches when certain photometric filter combinations are used \citep{Martocchia2018}. NGC 2173, with an age of 1.7 Gyr, is the youngest cluster discovered so far that exhibits a split RGB, which is a photometric signature of chemical abundance variations \citep{Kapse2022}. This is one example of a number of works (see \citealt{Salgado2022} for further examples) that show that the abundance patterns are not restricted to ancient globular clusters. However, since no spectroscopic analysis of clusters in the LMC younger than 2 Gyr have so far found evidence of chemical abundance spreads \citep{Mucciarelli2008}, we are still unsure of the age dependence for the occurrence of multiple populations in massive clusters. \par

In this work, we present a high-resolution study of elements including O, Na and Mg in NGC 1846, a LMC massive cluster with a mass of $\sim6\times10^{4} \Msol$ \citep{Song2019} and an age of 1.95 Gyr \citep{Goudfrooij2009}. This will allow a direct test of the 2 Gyr boundary for the age of massive star clusters exhibiting chemical abundance spreads. Compared to UV photometry, our approach allows the chemical abundance variations to be directly determined. In addition, since Na is unaffected by evolutionary mixing unlike in C and N, its abundance variation is a clear indicator of the presence of multiple populations in RGB stars \citep[e.g.,][]{Salgado2022}.\par

We present the observational material for the NGC 1846 RGB stars in Section~\ref{sec:observations}, and our photometric and spectroscopic analysis methods in Section~\ref{sec:photometric}. In Section~\ref{sec:abundance}, we present results of the abundance measurements based on both low and high-resolution spectroscopy. We also analyse the lack of any clear anti-correlation abundance signatures, and present statistical limits on the star-to-star abundance dispersion that may be present.

\section{Observations and Data Reduction}
\label{sec:observations}

\subsection{NGC 1846 RGB stars}
\label{sec:obs} 
The candidate cluster members were selected by \cite{MacKey2013} on the basis of CMD location, distance from the cluster centre and radial velocity.  As shown in their Figure 4, there is a well-defined group of candidate members centred on the known cluster velocity of $\sim$240 \kms (e.g., \citealt{Grocholski2006}) and lying within the 161 arcsec truncation radius \citep{Goudfrooij2009} for the cluster. In contrast, non-members in the same field have radial velocities in the wide range of 210--340 \kms. Furthermore, \cite{MacKey2013} calculated  membership probabilities for the candidates, finding $P_{mem} \gtrsim 99 \%$ in most cases; the lowest value is $P_{mem} = 92 \%$ for ACS-053 (see their Table 2).

\begin{figure}
	\includegraphics[width=\columnwidth]{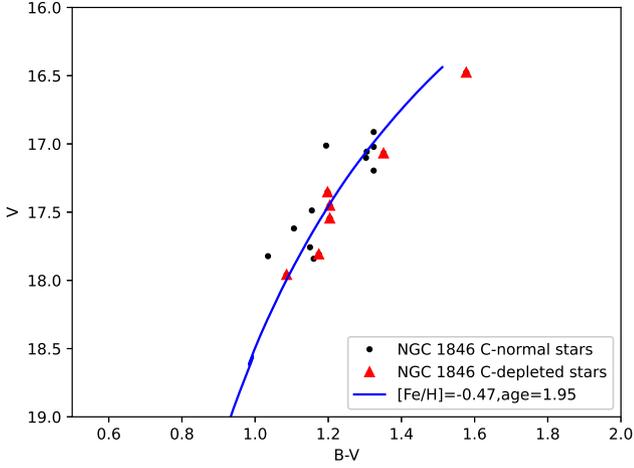}
    \caption{CMD of the NGC 1846 targets using B and V photometry from HST photometry presented by \protect\cite{MacKey2013}. ACS-043 and ACS-059 do not have B photometry, hence their values are omitted from the diagram. The red triangles indicate carbon depleted stars while the black points indicate carbon normal stars. This will be discussed in detail in Section 4.2. The Dartmouth RGB isochrone assumes literature values for the metallicity and age ([Fe/H] = $-0.47$, 1.95 Gyr; \protect\citealt{Goudfrooij2009}) has been included for reference.}
    \label{fig:CMD}
\end{figure}

Spectroscopic observations of 20 NGC 1846 RGB stars were obtained during three nights, 2008-11-30, 2008-12-01 and 2008-12-02 under ESO programme 082.D-0387 (PI: Mackey). These were obtained with the FLAMES instrument, which is a fibre-fed multi-object spectrograph mounted on the 8m ESO/VLT telescope. A total of 4 wavelength settings were employed (Table \ref{tab:res_table}), with three high-resolution settings (HR11, HR13, and HR14B) and one low-resolution one (LR02). The RGB stars observed, selected by \citet{MacKey2013} from their HST photometry, are sufficiently bright ($V \leq 19$) that sufficient S/N for high-precision abundance analysis is obtained. The CMD in Figure \ref{fig:CMD} shows the HST photometry from \citet{MacKey2013} for the NGC 1846 cluster members studied here. The data are well represented by a Dartmouth RGB isochrone assuming literature values for the cluster metallicity and age ([Fe/H] = $-0.47$, 1.95 Gyr; \citealt{Goudfrooij2009}). The lower panels in Figure 1 of \cite{MacKey2013} also show the location of additional stars in the cluster CMD.

We reduced the original raw FLAMES data with the standard ESO GIRAFFE pipeline \citep[esoreflex, version 2.16.7;][]{Blecha2000}. It performs all the basic reduction steps (bias removal, spectrum tracing, flat fielding, and wavelength calibration) together with sky and cosmic-ray subtraction. We did not account for the telluric lines in our spectra as they did not seem to interfere with the abundance measurements of our stars. In the last step, all exposures taken for each star are combined via simple addition to form the final spectra for analysis.
\footnote{As described in Mackey et al. (2013), the observations used the same optical fibre for the same star, and it was found that the change in barycentric correction during the period of observation is negligible.}

\begin{table}
	\centering
	\caption{ Observational setup for the spectroscopic data.}
	\label{tab:res_table}
	\begin{tabular}{lccr} 
		\hline
		Setting & Wavelength range (\AA) & Resolution\\
		\hline
		HR11 & 5597--5840 & 24200\\
		HR13 & 6120--6405 & 22500\\
		HR14B & 6383--6626 & 28800\\
		LR02 & 3964--4567 & 6000\\
		\hline
	\end{tabular}
\end{table}

\subsection{Abundance zero-point correction using Arcturus}

To obtain a reliable comparison of our measured abundances to the literature values, we must ensure that any systematic effects are accounted for. One way to do this is to apply our methods to the well-studied metal-poor Milky Way giant Arcturus (HD 124 897, $\alpha$ Boo) to obtain a zeropoint for our abundance scale. Since it has similar stellar parameters to the stars in the NGC 1846 sample ($\Teff = 4286$\,K, $\logg = 1.66$, $\FeH = -0.52$) \citep{Ramirez2011} and has also been used in other studies comparing the abundances of LMC stars \citep{VanderSwaelmenM.HillV.2013}, Arcturus is a good choice as a benchmark for our study. \par

We simulated Arcturus spectra in the four settings (LR02, HR11, HR13 and HR14B) by using the high resolution (R $\sim$ 150,000) \citep{Hinkle2000} spectral atlas of Arcturus. This was degraded according to the resolution required for each setting. The reference Arcturus abundances used are from \cite{Ramirez2011} and \cite{Worley2009}.

\section{Photometric and Spectroscopic analysis method}
\label{sec:photometric}

\subsection{Stellar parameters}

The surface gravities ($\logg$) for our NGC 1846 sample were derived canonically as shown in Equation~(\ref{eq:gravity}) below:

\begin{equation}
    \log\left(\frac{g}{g_\odot}\right)=\log\left(\frac{M}{M_\odot}\right)+4\log\left(\frac{T_\textrm{eff}}{T_\odot}\right)+0.4(M_\textrm{bol}-M_{\textrm{bol}\odot}).
	\label{eq:gravity}
\end{equation}

The steps we took were: Assuming masses of $\sim 1.5\Msol$ for the RGB stars, calculating the bolometric magnitudes using the absorption corrected V magnitudes from HST (E(B-V) = 0.036; \citealt{MacKey2013}), bolometric corrections described in \cite{Alonso1999} and assuming the LMC distance modulus to be $18.52 \pm 0.1$ mag \citep{Kovacs2000}. The solar bolometric reference value and effective temperature were taken to be 4.74 \citep{Mamajek2015} and 5770 K respectively. An initial effective temperature ($\Teff$) estimate was also provided using G--K$_s$ colour-temperature calibrations (using extinction corrected Gaia G and 2MASS K$_s$ magnitudes) from \cite{Casagrande2021}. 

$\Teff$ was then derived by interpolating our $\logg$ values onto a Dartmouth $\Teff$--$\logg$ isochrone \citep{Dotter2008} using reddening corrected V magnitudes from HST photometry, assuming literature values for metallicity and age ($\FeH = -0.47$, 1.95\,Gyr; \citealt{Goudfrooij2009}).\footnote{The values from Goudfrooij et al. (2009) were obtained from fitting the CMD derived from the HST photometry.} We tested the G--K$_s$ colour-temperature calibrations as mentioned earlier and found good agreement with our derived temperatures, where our mean bias and standard deviation are 20 K and 90 K respectively. This dispersion is similar to the median uncertainty in $\Teff$ from G--K$_s$ ($\sim120$ K). Given our methodology, it is not straightforward to determine the precision in stellar parameters. To estimate the precision for $\Teff$, we first interpolated the targets' V magnitude using the Dartmouth isochrone (from section 2.1) given a fixed B-V value. The new V magnitude was applied to recalculate the $\logg$ value, which was then used to determine a new $\Teff$ value. After which, we derived the offset between the recalculated and the actual $\Teff$ values, and its median absolute deviation was found to be 51 K.

Finally, the metallicities and $\vmic$ values for the NGC 1846 RGB stars were determined spectroscopically by fitting \ion{Fe}I \& \ion{Fe}{II} lines as described in the next section.

\begin{table}
	\centering
	\caption{Stellar parameters for the NGC 1846 RGB stars. Coordinates and photometry are detailed in Mackey et al. 2013.}
	\label{tab:stellar_param}
	\begin{tabular}{lcccc} 
		\hline
		Name & $\Teff$ (K) & $\logg$ & [Fe/H] & $\vmic$ (\kms)\\
		\hline
		ACS-001 & 3940 & 0.89 & -0.63 & 1.25\\
        ACS-013 & 4166 & 1.27 & -0.63 & 1.57\\
        ACS-017 & 4250 & 1.42 & -0.57 & 1.61\\
        ACS-025 & 4291 & 1.48 & -0.59 & 1.40\\
        ACS-030 & 4357 & 1.59 & -0.63 & 1.46\\
        ACS-036 & 4431 & 1.72 & -0.55 & 1.54\\
        ACS-043 & 4462 & 1.77 & -0.52 & 1.52\\
        ACS-046 & 4547 & 1.92 & -0.59 & 0.99\\
        ACS-047 & 4516 & 1.86 & -0.52 & 1.30\\
        ACS-053 & 4534 & 1.89 & -0.68 & 1.47\\
        ACS-059 & 4588 & 1.99 & -0.50 & 1.06\\
        ACS-066 & 4638 & 2.08 & -0.59 & 1.38\\
        ACS-080 & 4212 & 1.35 & -0.62 & 1.43\\
        ACS-081 & 4216 & 1.36 & -0.60 & 1.50\\
        ACS-082 & 4231 & 1.38 & -0.67 & 1.45\\
        ACS-085 & 4236 & 1.39 & -0.61 & 1.52\\
        ACS-090 & 4395 & 1.66 & -0.60 & 1.25\\
        ACS-092 & 4410 & 1.68 & -0.58 & 1.32\\
        ACS-102 & 4540 & 1.90 & -0.61 & 1.26\\
        ACS-112 & 4468 & 1.78 & -0.65 & 1.04\\
		\hline
	\end{tabular}
\end{table}

\subsection{Abundance analysis}
\subsubsection{High-resolution Abundance analysis}

For our spectroscopic analysis, we used the spectrum synthesis code Spectroscopy Made Easy (SME) (version 536) \citep{Piskunov2017} and 1D MARCS model atmospheres \citep{Gustafsson2008}. We implemented NLTE corrections using pre-tabulated grids of departure coefficients for O, Na, Mg, Si, Ca, Ba \citep{Amarsi2020} and Fe \citep{Amarsi2022}. The rest of the elemental abundances were computed assuming LTE. We used atomic and molecular line data from VALD3 \citep{Ryabchikova2015}. 

Continuum and line masks were defined mostly by hand, by inspecting a number of spectra to avoid features that appeared too blended or influenced by telluric contamination. The continuum was fitted by dividing the observations by a synthetic spectrum, and fitting a straight line in selected continuum windows. This was done for segments of $\sim50$ \AA\/ in length. The Fe lines that were used to estimate [Fe/H] were carefully chosen depending on how well the synthetic spectra fit the observed ones. Since the Fe lines found in the HR13 setting were found to be of the best quality amongst all the high-resolution settings, we decided to use the $\vmic$ values derived from that setting. As SME performs a global $\chi^2$ fit between synthetic and observed spectra, we determined $\FeH$ individually from each setting and adopted their average as our final metallicity. We chose this approach on the basis that the other elements are located in the various spectrograph settings, and therefore the average Fe value is more representative than using the Fe value from one particular setting. For other elements, we implemented a similar approach to that of Fe. However, a key difference is that in cases where an element could be measured in several settings, we picked the setting that yielded the smallest formal errors.

\subsubsection{Low-resolution Abundance analysis}

A grid of synthetic spectra were used for our low-res spectroscopic analysis to measure carbon. 
Spectra were computed as described in \cite{Nordlander2019}, using a grid of MARCS model atmospheres \citep{Gustafsson2008}, the synthesis code TurboSpectrum (v15.1; \citealt{Plez2012}), atomic line data from VALD3 \citep{Ryabchikova2015} and with molecular data for CH \citep{Masseron2014} and CN \citep{Brooke2014,Sneden2014} as well as numerous other molecules. We adopted a metallicity-dependent alpha enhancement based on typical values in the Milky Way's disk, $\rm[\alpha/Fe] = -0.4[Fe/H]$, that matches the adopted model atmosphere grid (i.e. $\rm [alpha/Fe] \approx 0.2$ at $\rm [Fe/H] \approx -0.5$), and computed spectra over a range of $\vmic$, $\rm [C/Fe]$ and $\rm[N/Fe]$ values for each model atmosphere in the grid. 

The same stellar parameters were used as for the high-res analysis. We fitted spectra using a $\chi^2$ minimisation, and used a maximum likelihood analysis to ensure detections were significant above the noise level.
The continuum and molecular absorption regions were carefully chosen, with the latter being the CH G-band found at $\sim$4300~\AA. As for nitrogen, we were not able to obtain any reliable abundance measurements as tests on the CN band (4120--4216\,\AA) showed that it was not possible to provide meaningful constraints on the N abundance. 	

\subsection{Error analysis}
We estimated total uncertainties for the abundance measurements by combining statistical and systematic errors.
We adopted statistical error estimates from the $\chi^2$ minimisation routine for our high and low-res abundances measurements. Both use the Levenberg-Marquardt $\chi^2$ optimisation and we take $\sigma^2$ from the diagonal of the covariance matrix.

The systematic errors are based on uncertainties in the stellar parameters. Due to the high precision in the V magnitudes, the scatter in $\Teff$ itself is minimal. Hence, we assume the correlated error between $\Teff$ and $\logg$ to be the leading error term, where $\Teff$ and $\logg$ vary in tandem. Taking the error in $\Teff$ to be 50 K as described in section 3.1, we find that a shift of 50\,K in $\Teff$ along the isochrone corresponds to a 0.1\,dex change in $\logg$. We note that there is an additional error term in the $\Teff$ and $\logg$ scales, due to uncertainties in the adopted reddening, distance modulus, mass and metallicity. But these are comparable to the precision that we derived earlier ($\sim$0.1 dex for $\logg$ with a correlated 50\,K error in $\Teff$), and would have a similar impact on all stars, leading to negligible star-to-star abundance differences. The $\vmic$ and [Fe/H] errors were calculated by adopting the standard deviations of the $\vmic$ (0.2\,\kms) and [Fe/H] measurements (0.05\,dex) respectively. In comparison, we find that perturbing $\Teff$ and $\logg$ leads to relatively minor changes in $\vmic$ (0.02\,\kms) and [Fe/H] (0.01\,dex). These uncertainties were then used to perturb the stellar parameters and compute the change in abundance measurements, which were combined in quadrature to compute the total systematic uncertainty. Finally, the total error was calculated by simply combining the statistical and systematic errors in quadrature.

\subsection{Obtaining final abundance measurements}

We provide both raw and calibrated abundances. The latter is to account for the systematic trend between the abundance and $\Teff$ that occurs for some elements. This is done by fitting a straight line to the measurements and removing the slope while retaining the mean. The slopes are provided in appendix \ref{tab:slope} in units of dex/1000K. Finally, offsets were applied to our measured abundances by using the Arcturus abundance measurements to obtain the zero-point corrections for each element.  Unless otherwise specified, we will use the calibrated and zero-point corrected abundances in the rest of our analysis.

\begin{table}
	\centering
	\caption{The chemical composition estimated for Arcturus in this work, reference abundances by \citep{Ramirez2011} (1) and \citep{Worley2009} (2), and the offsets we applied to match the literature abundance scale.}
	\label{tab:Arcturus}
	\begin{tabular}{lrrrr} 
		\hline
		Element & This work & Literature & Offset & Ref\\
		\hline
		$\rm [C/Fe]$	& $-0.08$ & 0.42 & $-0.50$ & 1\\
		$\rm [O/Fe]$	& 0.50	& 0.50 & 0.00 & 1\\
		$\rm [Na/Fe]$	& 0.33	& 0.11	& 0.22 & 1\\
		$\rm [Mg/Fe]$	& 0.68	& 0.37	& 0.31 & 1\\
		$\rm [Si/Fe]$	& 0.12	& 0.33	& $-0.21$ & 1\\
		$\rm [Ca/Fe]$	& 0.09	& 0.11	& $-0.02$ & 1\\
		$\rm [Ti/Fe]$	& 0.26	& 0.27	& $-0.01$ & 1\\
		$\rm [Fe/H]$    & $-0.73$ & $-0.52$	& $-0.21$ & 1\\
		$\rm [V/Fe]$    & 0.25 & 0.20	& 0.05 & 1\\
		$\rm [Zr/Fe]$	& 0.13 & 0.01 & 0.12 & 2\\
		$\rm [Ba/Fe]$	& 0.06	& $-0.19$ & 0.25 & 2\\
		$\rm [Eu/Fe]$	& 0.36	& 0.36	& 0.00 & 2\\
		\hline
	\end{tabular}
\end{table}

\section{Constraining the presence of Multiple Populations in NGC 1846} 
\label{sec:abundance}

In this section, we will present the results showing the lack of evidence for chemical inhomogeneities in Na and O in NGC 1846, indicating there is no evidence for multiple populations present in the cluster. We also present the upper limits for the star-to-star intrinsic abundance scatter for all elements measured.

\subsection{Cluster mean abundances and star-to-star variations} 

\begin{figure}
	\includegraphics[width=\columnwidth]{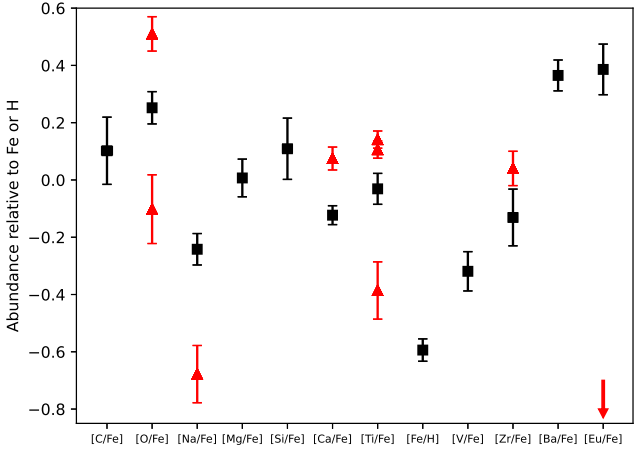}
    \caption{Results of the abundance analysis, represented as a scatter plot, where the black squares indicate the median abundance value and the error bars indicate the median absolute deviation. The red triangles represent outlier points together with their total measurement errors. The outlier for Eu has $\rm [Eu/Fe] = -2 \pm 5$ and is indicated by the red arrow; this is not a genuine detection as indicated by the extremely large error estimate. The outlier stars are: O (ACS-059, ACS-066), Na (ACS-053), Ca (ACS-046), Ti (ACS-001, ACS-066, ACS-112), Zr (ACS-001) and Eu (ACS-059).}
    \label{fig:box}
\end{figure}

Figure~\ref{fig:box} and Table~\ref{tab:mean} indicate the observed star-to-star spreads and mean abundances respectively for all measured elements. To determine the intrinsic scatter of the abundance measurements that may be hiding beneath the combined errors, we used a Markov Chain Monte Carlo (MCMC) code \citep{Foreman-Mackey2013} to determine the maximum likelihood value of the intrinsic spread in abundance ratio [X/Fe] for each element. This was done by simulating the total spread in our measurements which combines the statistical, systematic and intrinsic dispersion in quadrature. A uniform prior was adopted. Contour plots were also generated for easier visualization of the data.

Outliers were removed from our data by applying statistical thresholds, which were either if the residual from a straight line fit exceeds 3 $\sigma$ of the residual sample, or if the difference between the abundance and mean exceeds three times of the combined measurement error. We do not know the actual cause of the outliers even after inspecting the respective spectral regions, and they do not appear to be due to systematic errors in stellar parameters. We therefore choose a conservative approach of reporting the status of the outliers, but do not speculate further on the cause.

We have found that the star-to-star dispersion is comparable to the measurement uncertainties for most of the elements, which means that the intrinsic star-to-star spreads for these elements are small (median $\sigma_{\rm int}$ $\leq0.04$) as shown in Table \ref{tab:spread}. The only exceptions are for Zr, Si and C. For Zr and Si, we can attribute this to the relatively large measurement uncertainties based on the standard deviation of the intrinsic abundance
dispersion ($\geq 0.03$), and note that is not unusual for a false positive at the 2 sigma confidence level to arise in a sample of 20 tests. C will be further elaborated in section 4.3. 

\begin{table}
	\centering
	\caption{The mean calibrated abundances and the median absolute deviation (MAD) for all elements, plus the spectral regions in which the measured lines occur. The standard error of the mean is also given; systematic errors are not accounted for in these values.}
	\label{tab:mean}
	\begin{tabular}{lrrl} 
		\hline
		Element & Mean & MAD & Spectral region(s)\\
		\hline
		$\rm [C/Fe]$	& $0.10 \pm0.01$ & 0.12 & LR02\\
		$\rm [O/Fe]$	& $0.25 \pm0.02$	& 0.06 & HR13\\
		$\rm [Na/Fe]$	& $-0.24 \pm0.02$ & 0.05 & HR13\\
		$\rm [Mg/Fe]$	& $0.01 \pm0.02$	& 0.07 & HR13\\
		$\rm [Si/Fe]$	& $0.11 \pm0.02$	& 0.11 & HR13\\
		$\rm [Ca/Fe]$	& $-0.12 \pm0.01$	& 0.03 & HR14B\\
		$\rm [Ti/Fe]$	& $-0.03 \pm0.01$	& 0.05 & HR14B\\
		$\rm [Fe/H]$    & $-0.59\pm0.01$ & 0.04 & HR11, HR13, HR14B\\
		$\rm [V/Fe]$    & $-0.32 \pm0.02$ & 0.07 & HR13\\
		$\rm [Zr/Fe]$	& $-0.13 \pm0.02$ & 0.10 & HR13\\
		$\rm [Ba/Fe]$	& $0.37 \pm0.02$	& 0.05 & HR13\\
		$\rm [Eu/Fe]$	& $0.39 \pm0.03$	& 0.09 & HR14B\\
		\hline
	\end{tabular}
\end{table}

\begin{table}
	\centering
	\caption{Columns showing the median intrinsic dispersion ($\sigma_\text{int}$), number of stars included in the sample, twice the standard deviation of the intrinsic dispersion (2 SD($\sigma_\text{int}$)), and 95\,\% confidence limit on the maximum $\sigma_\text{int}$ for each element measured.}
	\label{tab:spread}
	\begin{tabular}{lccccc} 
		\hline
		Element & $N_\text{stars}$ & Median $\sigma_\text{int}$ & 2 SD($\sigma_\text{int}$) & 95\,\% limit on $\sigma_\text{int}$ \\
		\hline
		$\rm [C/Fe]$ & 20   & 0.14	& 0.05	& $\leq0.19$\\
		$\rm [O/Fe]$ & 18	& 0.04	& 0.05	& $\leq0.09$\\
        $\rm [Na/Fe]$ & 19	& 0.02	& 0.04	& $\leq0.07$\\
        $\rm [Mg/Fe]$ & 20	& 0.02	& 0.03	& $\leq0.06$\\
        $\rm [Si/Fe]$ & 20	& 0.07	& 0.08	& $\leq0.13$\\
        $\rm [Ca/Fe]$ & 19	& 0.02	& 0.03	& $\leq0.05$\\
        $\rm [Ti/Fe]$ & 17	& 0.04	& 0.04	& $\leq0.07$\\
        $\rm [Fe/H]$ & 20	& 0.02	& 0.03	& $\leq0.05$\\
        $\rm [V/Fe]$ & 20	& 0.02	& 0.04	& $\leq0.07$\\
        $\rm [Zr/Fe]$ & 19	& 0.05	& 0.07	& $\leq0.12$\\
        $\rm [Ba/Fe]$ & 20	& 0.02	& 0.03	& $\leq0.05$\\
        $\rm [Eu/Fe]$ & 19	& 0.04	& 0.06	& $\leq0.10$\\
        
		\hline
	\end{tabular}
\end{table}

\begin{figure*}
	\includegraphics[width=\columnwidth]{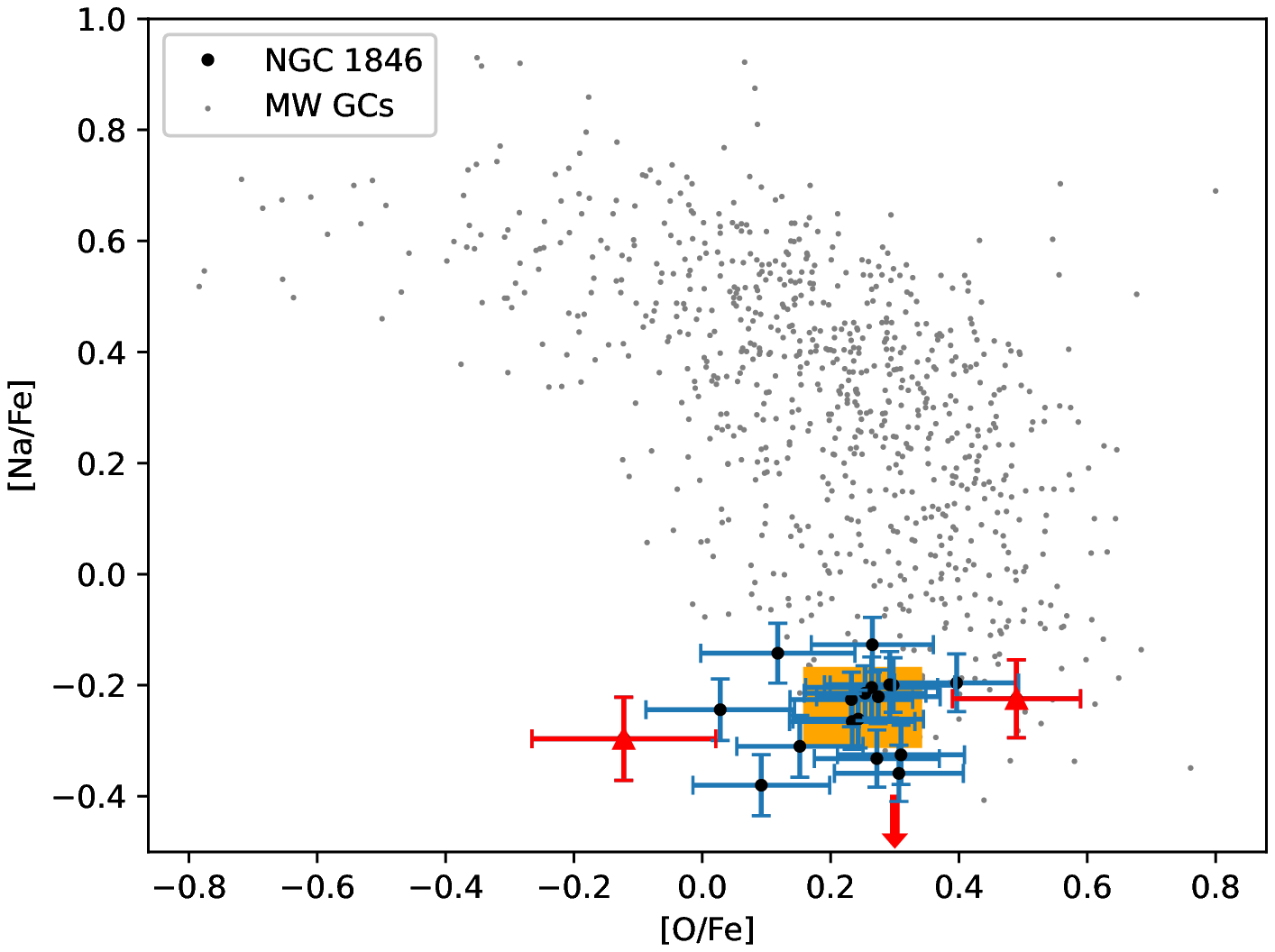}
	\includegraphics[width=\columnwidth]{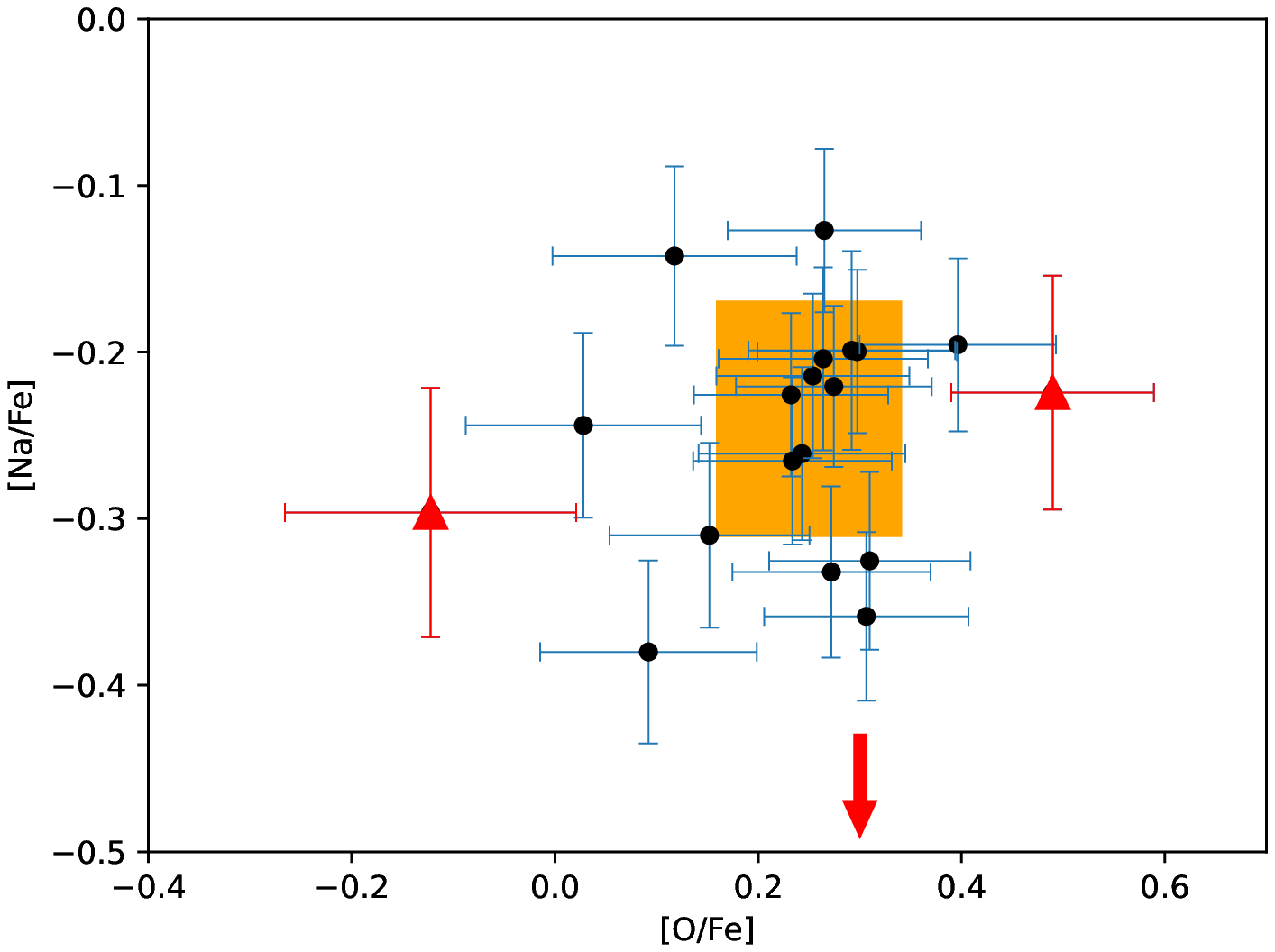}
    \caption{The left plot shows the comparison of the Na and O abundances in the NGC 1846 stars with literature globular cluster values \protect\citep{Carretta2009}. Red triangles indicate the outlier points for O, while the red arrow indicates the outlier point for Na. Error bars denote the combined random and systematic uncertainties associated with each measurement. The orange shaded box covers the central 95 \% of the likelihood distribution for the intrinsic spread in Na and O. The right plot shows a zoomed in version of the left plot.}
    \label{fig:NaO}
\end{figure*}

The abundances of Na and O obtained from our high-res spectroscopy show no sign of any anti-correlation, as shown in Figure~\ref{fig:NaO}, which also shows literature data for Milky Way Globular Clusters from \cite{Carretta2009}. The formal maximum-likelihood analysis constrains the spreads in O and Na to be $\sigma_{\rm int} \le 0.09$ and $\le 0.07$\,dex at 95\,\% confidence, respectively, as shown in Table \ref{tab:spread}. Corner plots for Na, Mg and O are included in Appendix \ref{fig:corners}. This confirms that there is no evidence for MPs in NGC 1846.

\subsection{Carbon analysis}

\begin{figure}
	\includegraphics[width=\columnwidth]{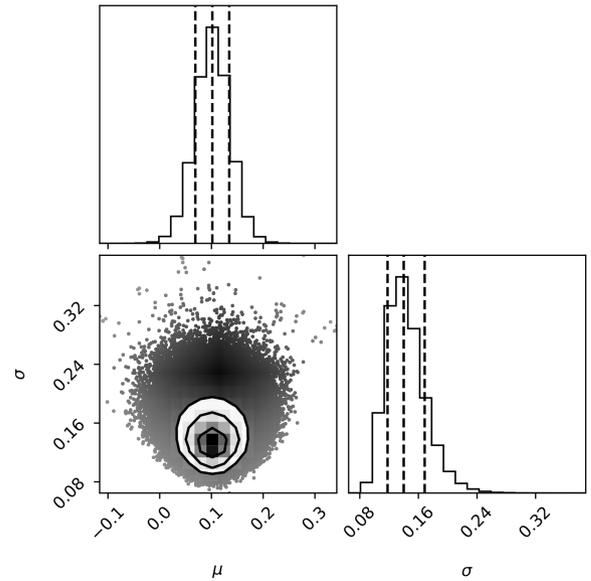}
    \caption{Corner plot showing the mean abundance ($\mu$) and abundance dispersion ($\sigma_\text{int}$) of carbon in NGC 1846. The dashed lines indicate the 25, 50 and 75 percentiles of the intrinsic spread of carbon. The median spread of carbon is $0.140 \pm 0.051$.}
    \label{fig:corner}
\end{figure}

Our maximum-likelihood analysis indicates that C is the only element that exhibits a robust non-zero star-to-star spread, as shown in the corner plot in Figure \ref{fig:corner}. While some other elements in the scatter plot (Figure \ref{fig:box}) apparently show a comparable spread, their measurement uncertainties are commensurately larger (Refer to Table~\ref{tab:complete} in the appendix).

\begin{figure}
	\includegraphics[width=\columnwidth]{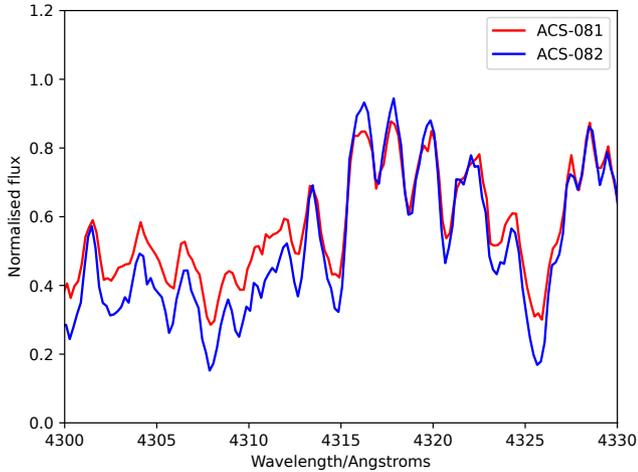}
    \caption{G-band spectral segment showing the difference in CH feature strength for the NGC 1846 RGB stars ACS-081 ($\Teff = 4216$\,K) and ACS-082 ($\Teff = 4231$\,K). The determined [C/Fe] values from synthetic spectra fits are --0.25 and +0.20, respectively.}
    \label{fig:C_com}
\end{figure}

To support our finding, we show in Figure~\ref{fig:C_com} a spectral segment in the vicinity of the CH G-band for two NGC 1846 RGB stars (ACS-081 \& ACS-082) that have similar stellar parameters but which have significantly different [C/Fe] (values of --0.25 vs 0.20). 

\begin{figure}
	\includegraphics[width=\columnwidth]{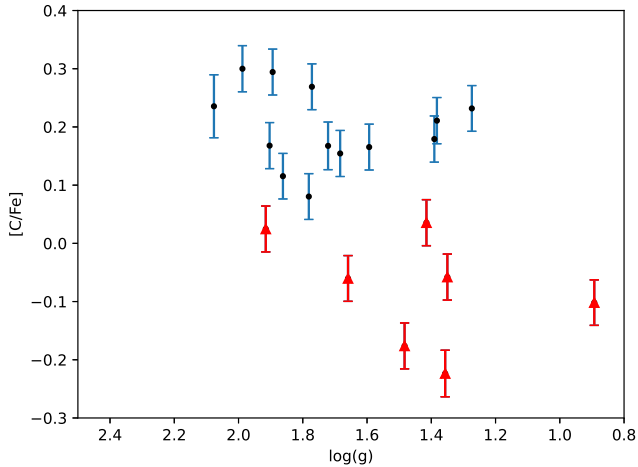}
    \caption{[C/Fe] vs log(g) plot for the NGC 1846 RGB sample. The red triangles indicate carbon depleted stars while the black points indicate carbon normal stars. The error bars show the combined random and systematic errors of the [C/Fe] abundance measurements for every star.}
    \label{fig:C_logg}
\end{figure}

We also observe a decreasing C abundance with decreasing log(g) based on Figure \ref{fig:C_logg}, which we interpret as a signature of evolutionary mixing. Potential mixing processes include thermohaline mixing and meridional circulation that occur as stars ascend the RGB, bringing up material from deeper layers that has been processed via thermonuclear CNO burning that converts C into N \citep{Karakas2014}. However, the spread in our measurements indicates that the degree of mixing varies from star to star, even at similar log(g). This suggests an additional parameter is involved in governing the mixing process.

It is not possible using our available photometry to distinguish whether a star belongs to the RGB or the AGB.  Tests with a MIST isochrone indicate that the RGB:AGB ratio in our sample is likely to be 3:1 with an even distribution as a function of $\logg$. It is therefore possible that our most C-depleted stars are AGB, while the rest are RGB. We have marked the most C-depleted stars in Fig.~\ref{fig:C_logg} and in the CMD in Fig.~\ref{fig:CMD}, which indicates that the C-depleted stars have photometry that is fully compatible with the RGB isochrone. 
We note in particular that at the luminosities of our stars, AGB stars are double-shell source stars but have not yet reached the regime of third dredge-up where surface carbon abundances increase. Instead, the surface carbon in these stars is expected to be similarly depleted as for our most luminous RGB stars that are approaching the RGB tip.

One possibility is that differences in the rotation velocities of the stars can lead to varying amounts of mixing. \cite{Chaname2005} predicts that for old field giants, a 30\,\kms\ difference in initial rotation can lead to 0.5 dex variations in carbon abundance, with stronger rotation leading to stronger depletion, which is similar to what we observe in Figure \ref{fig:C_logg}. Moreover, as mentioned in the introduction, \cite{Kamann2020} have shown that the stars in the eMSTO region of NGC 1846 display varying surface rotation rates (60--180 \kms). Inspection of our highest resolution spectra, however, show no indication of surface rotation higher than 10 \kms\ in any of our RGB stars. Nevertheless, even though rotation stops on the surface as the star evolves from MS to RGB, it could continue in the core, leading to mixing effects in the upper layers of the star.
We note that while \cite{Chaname2005} only predicted significant depletion of carbon in their rotating models, modern calculations do so even in non-rotating models. As discussed by \citet{Karakas2014}, it is not necessarily true that diffusion coefficients from thermohaline mixing and rotation simply add together -- it is possible that rotation actually inhibits thermohaline mixing, and so the effect of rotation on RGB surface abundances may be the opposite of what is discussed above.

Therefore, our results show that it is possible to have a significant intrinsic spread of surface carbon abundance in a cluster without any star-to-star variations in most of the other element abundances (including O and Na that are the characteristic signature of MPs in ancient GCs). However, we cannot rule out N variations in our sample. Since hydrogen burning conserves the sum C+N, these stars must also have varying [N/Fe]. These variations may be what has been detected in past photometric surveys of RGB stars. Measuring nitrogen abundances for our sample would thus be an important follow-up project. 

\section{NGC 1846 in the LMC context}

\begin{figure*}
	\includegraphics[width=\columnwidth]{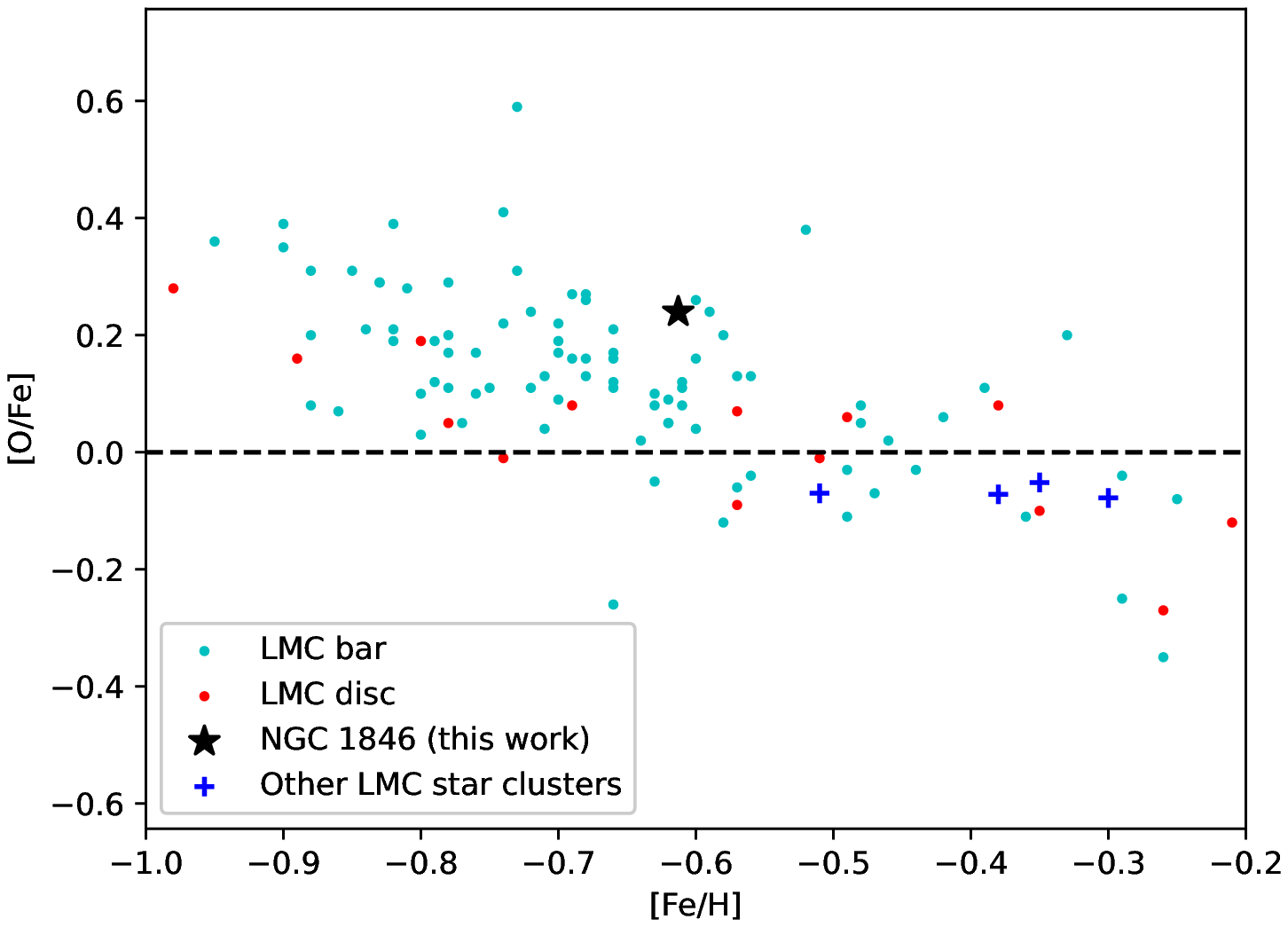}
	\includegraphics[width=\columnwidth]{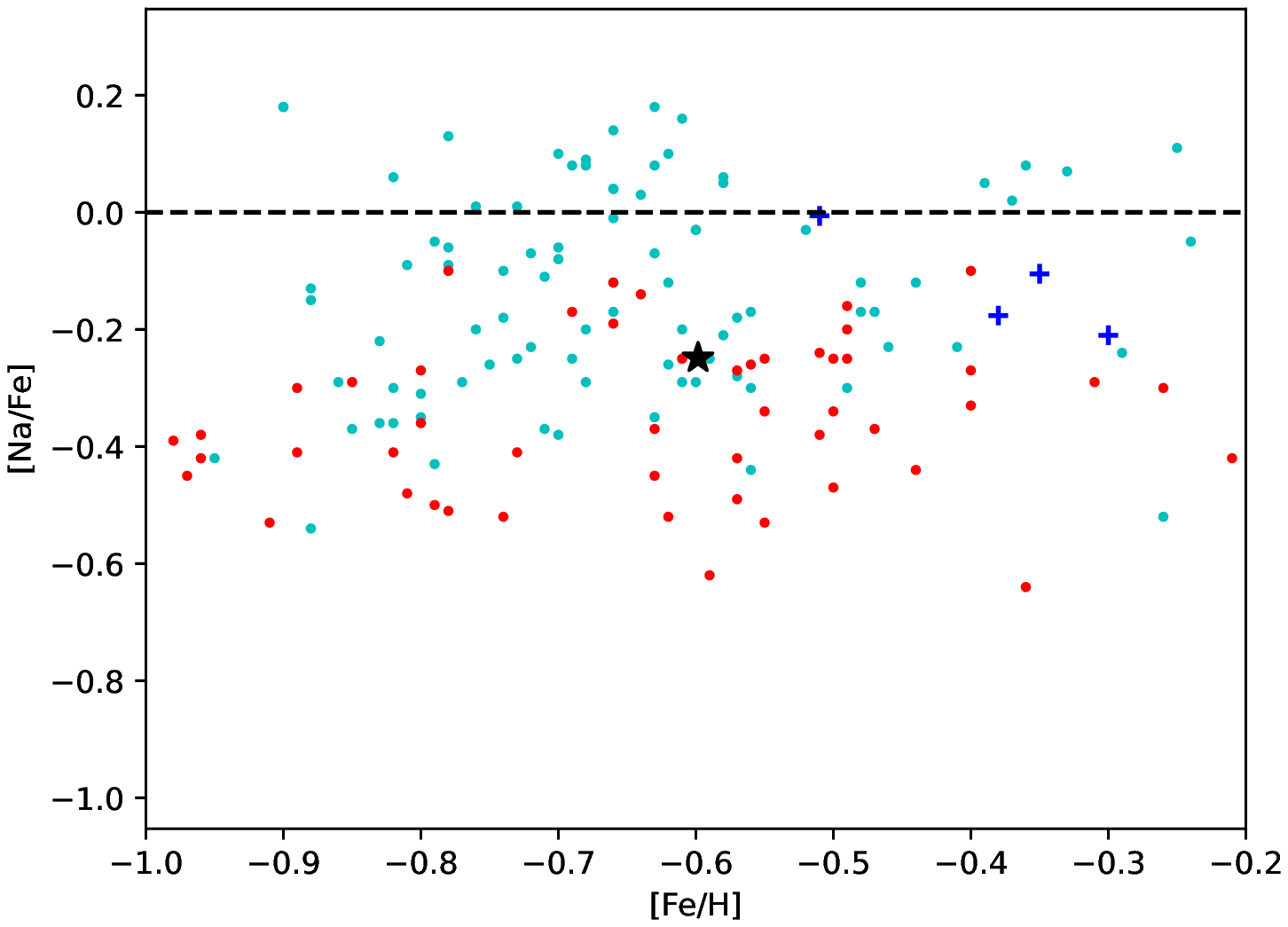}
	\includegraphics[width=\columnwidth]{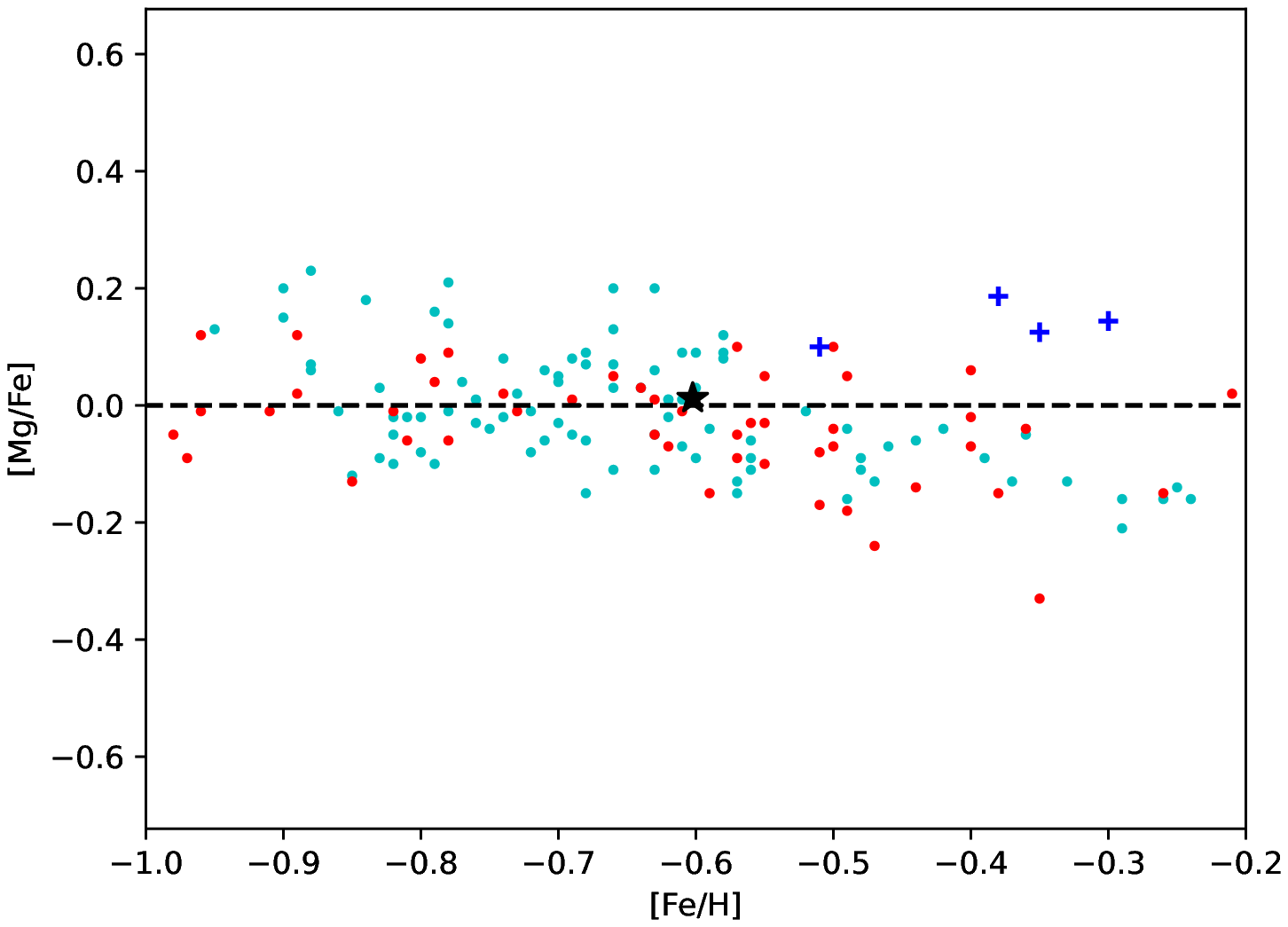}
	\includegraphics[width=\columnwidth]{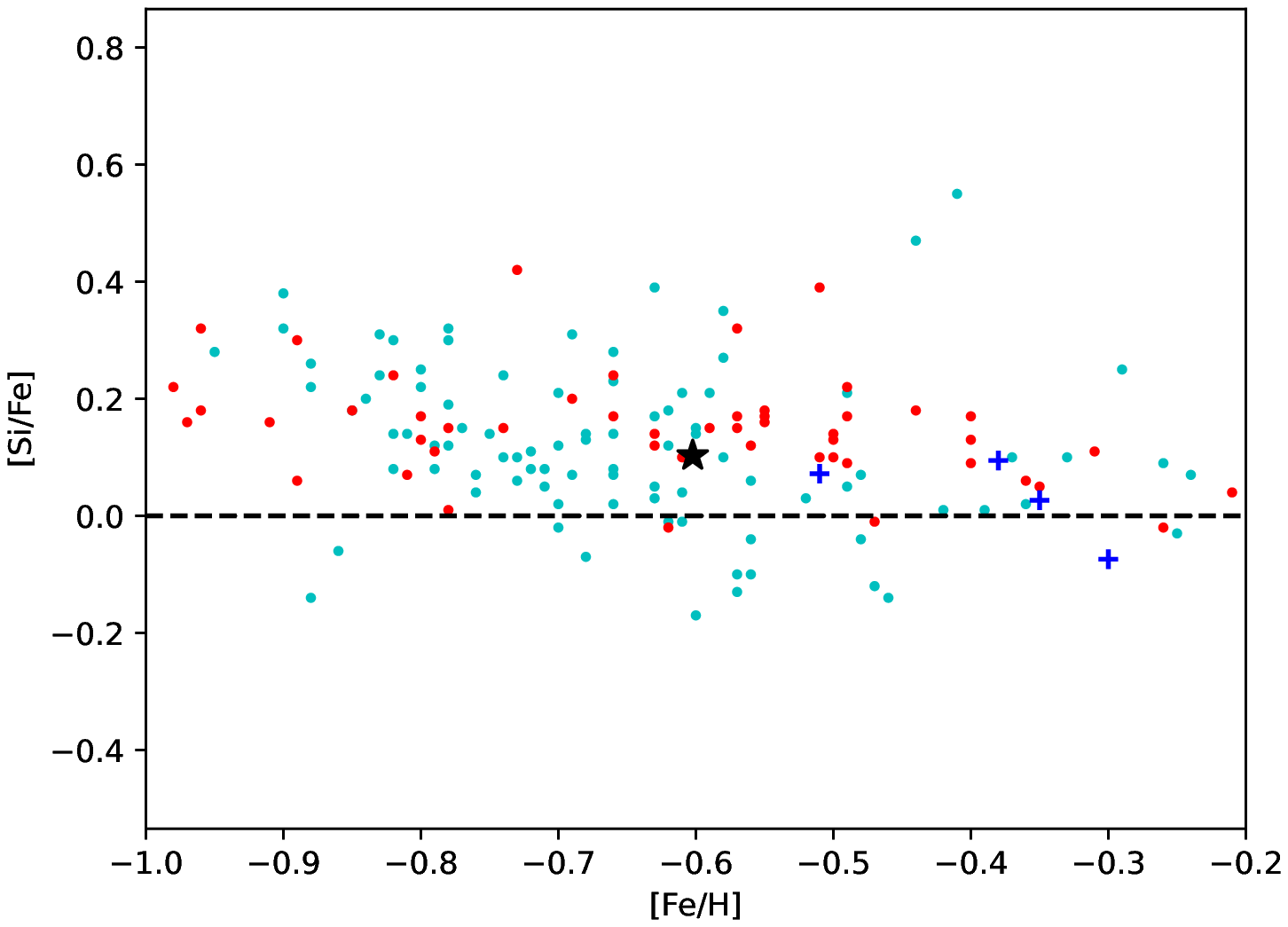}
	\includegraphics[width=\columnwidth]{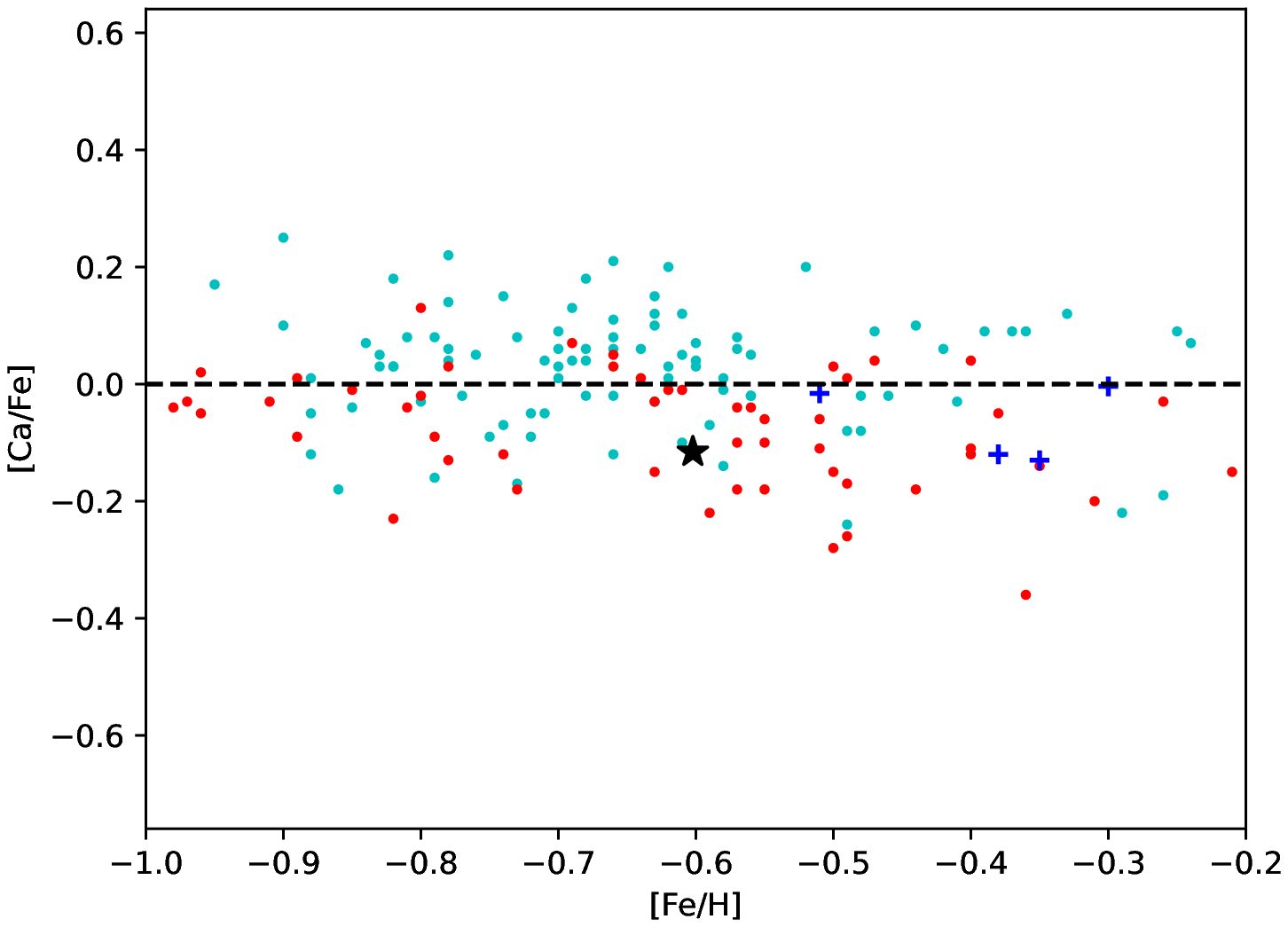}
	\includegraphics[width=\columnwidth]{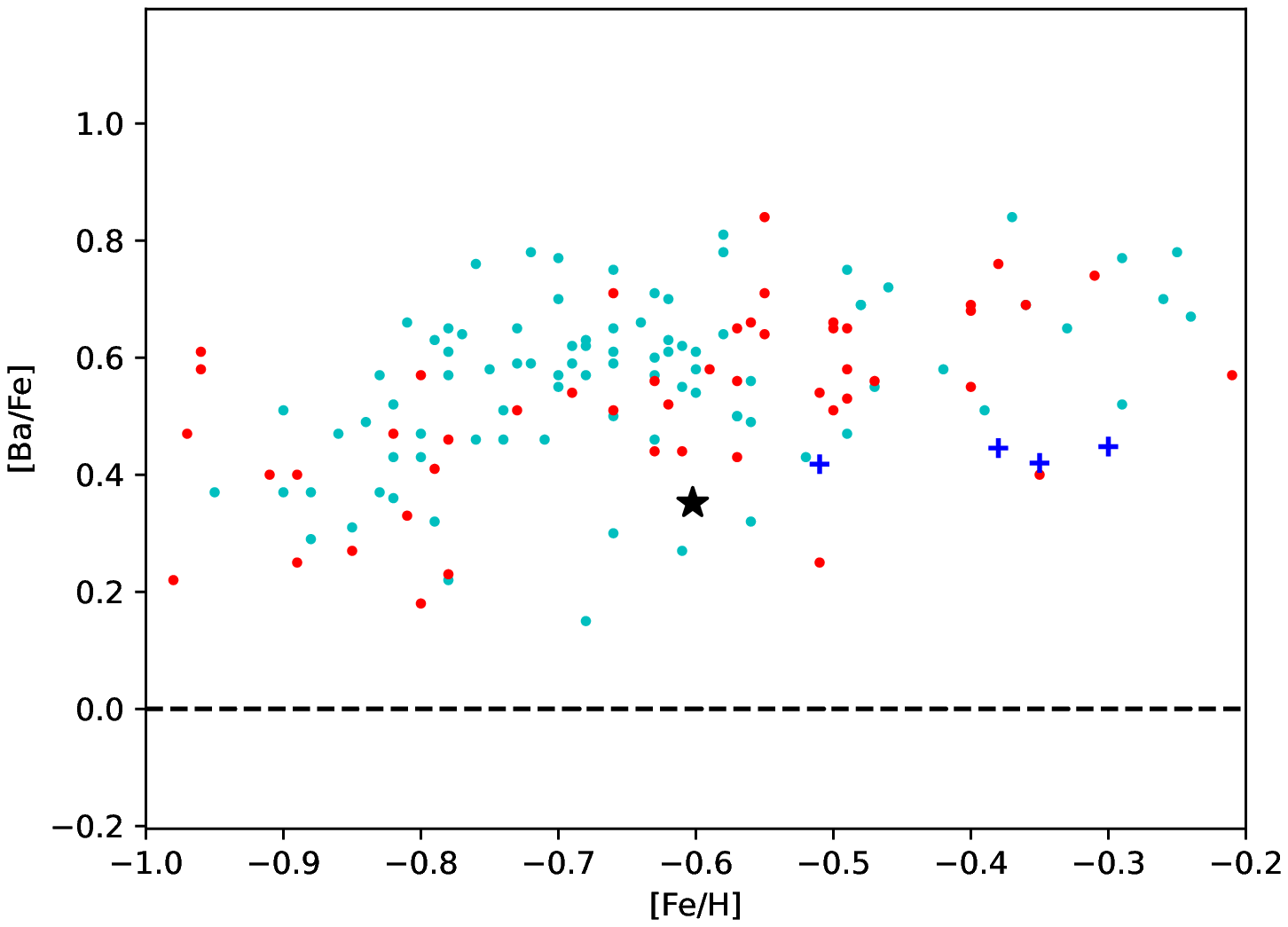}
    \caption{Comparison of our abundances for stars in NGC 1846 (black star) to past studies of LMC field stars (bar - cyan and inner disc - red; \protect\citealt{VanderSwaelmenM.HillV.2013}) and for other LMC $\sim$2 Gyr old star clusters (blue crosses; \protect\citealt{Mucciarelli2008}). Corrections for non-local thermodynamic equilibrium effects were included for these elements. The black dashed line shows the solar abundance level.}
    \label{fig:Na_com}
\end{figure*}

\begin{figure*}
	\includegraphics[width=\columnwidth]{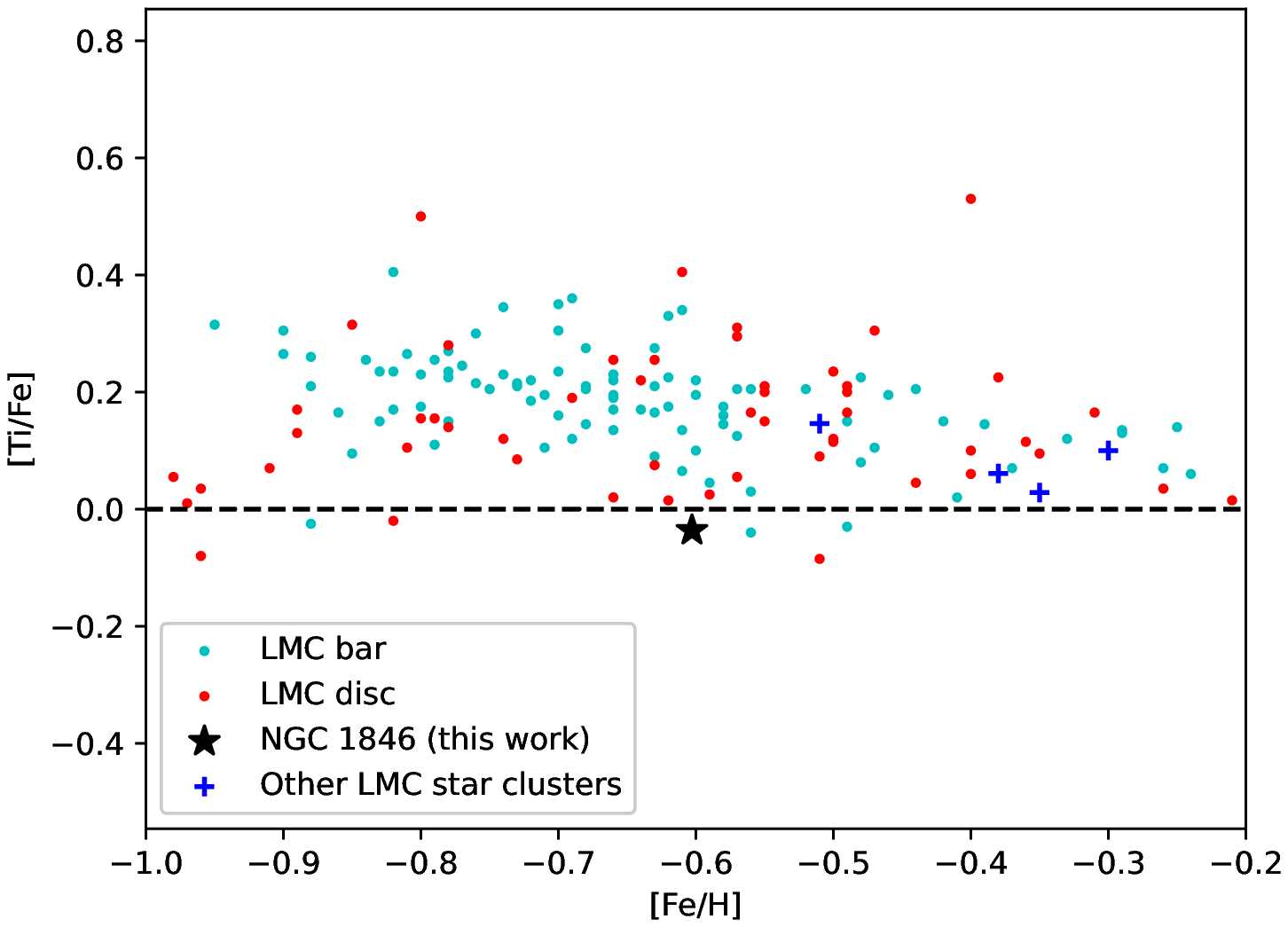}
	\includegraphics[width=\columnwidth]{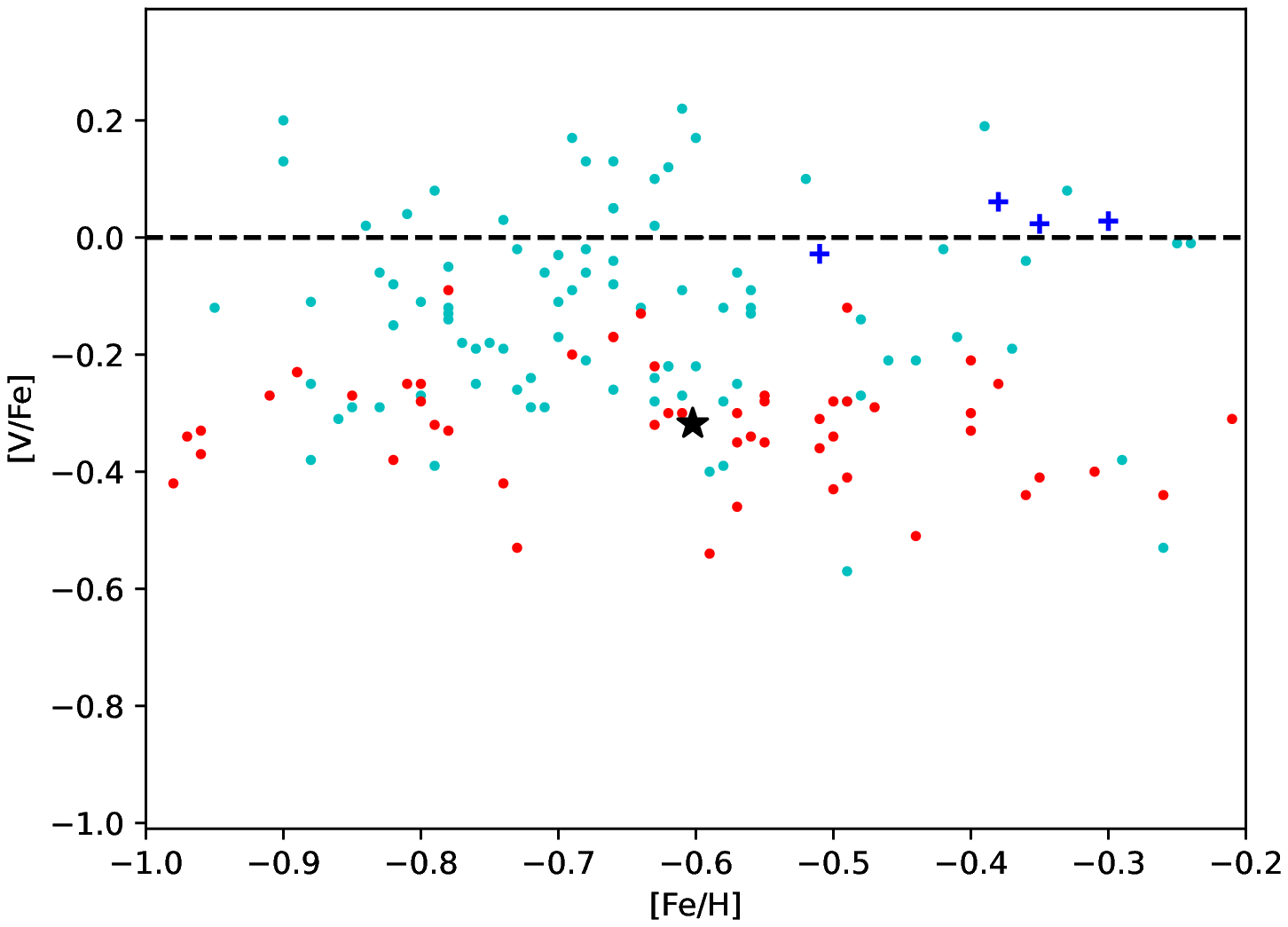}
	\includegraphics[width=\columnwidth]{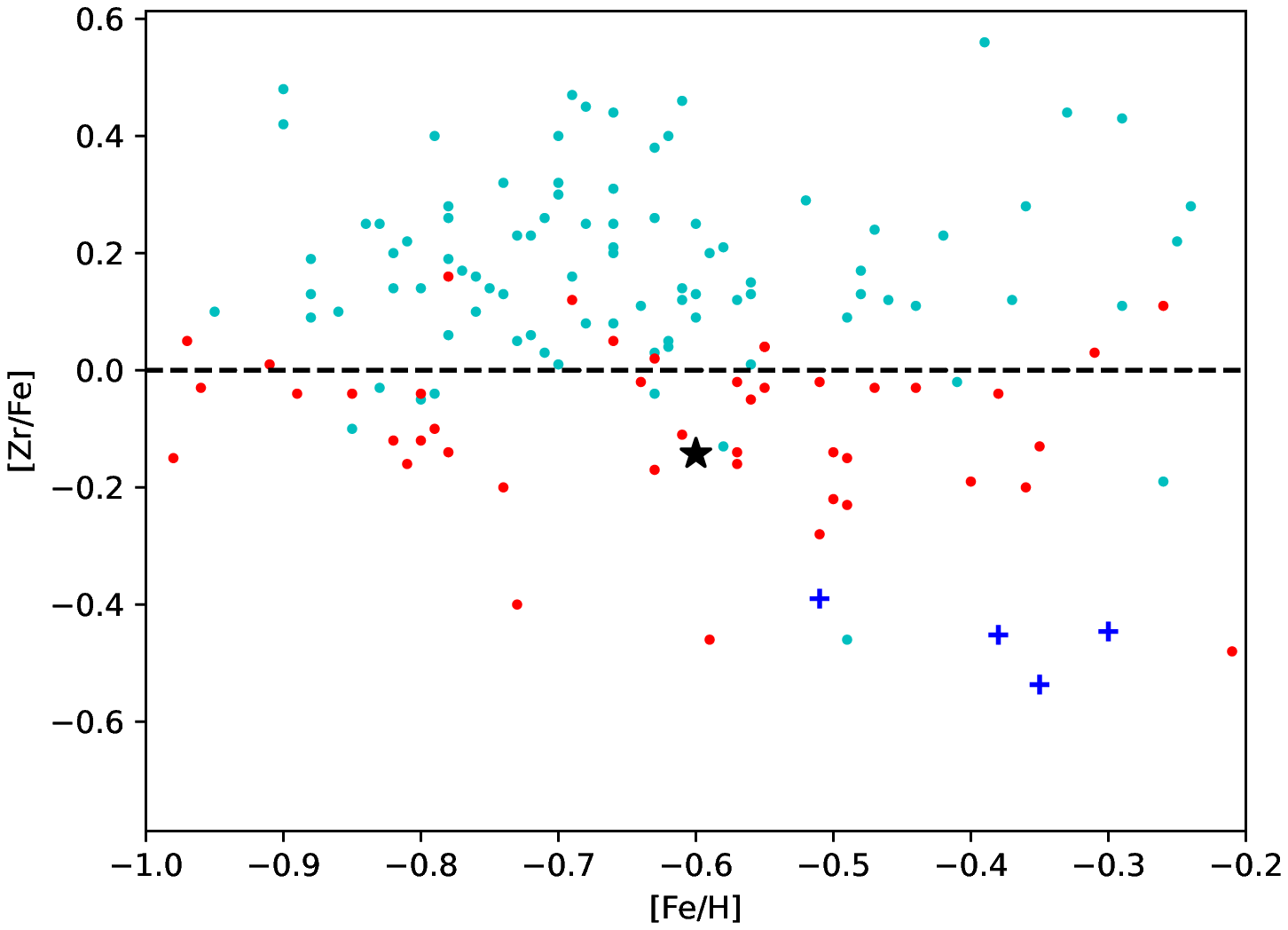}
	\includegraphics[width=\columnwidth]{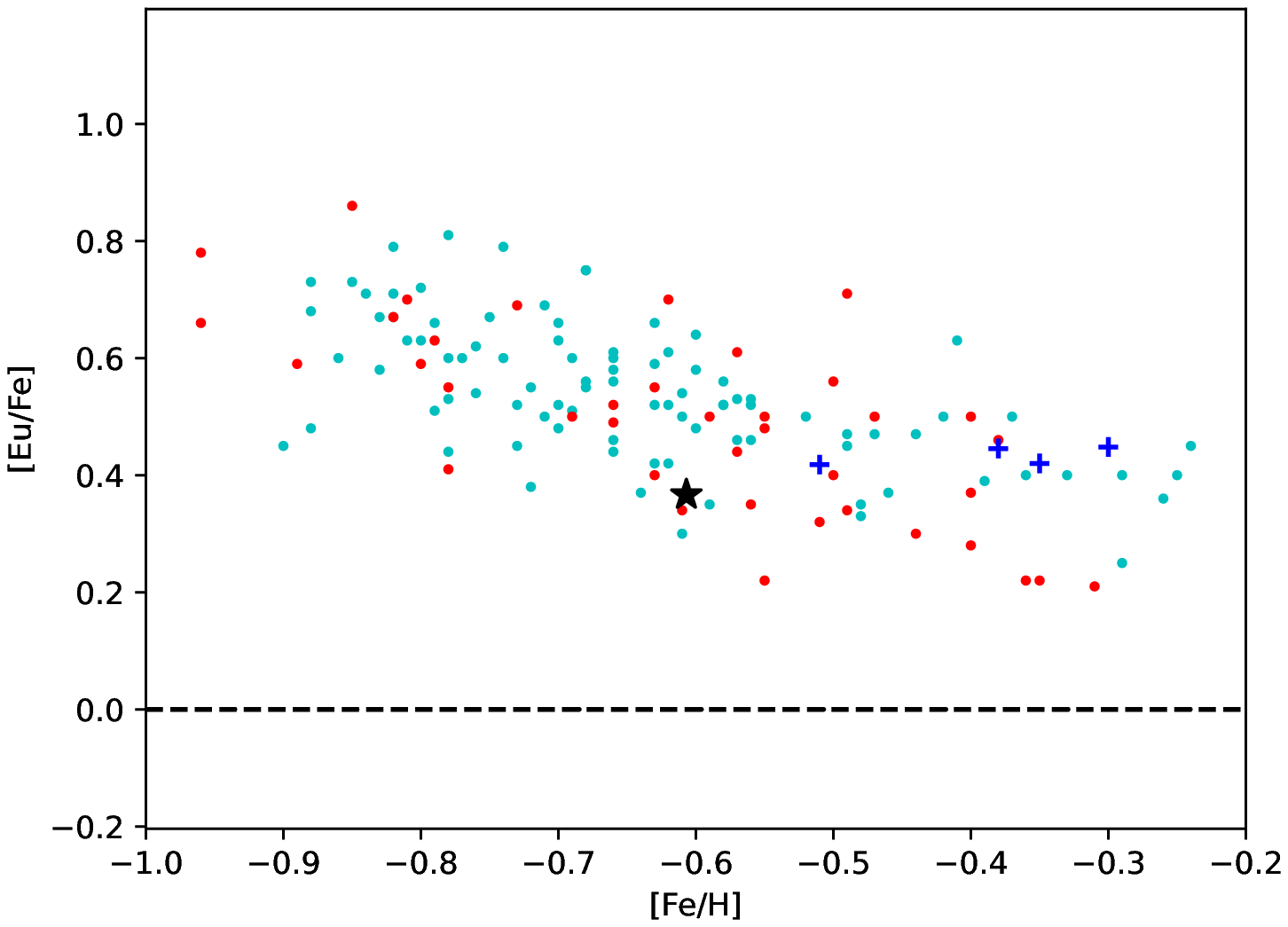}
    \caption{Comparison of our abundances for stars in NGC 1846 (black star) to past studies of LMC field stars (bar - cyan and inner disc - red; \protect\citealt{VanderSwaelmenM.HillV.2013}) and for other LMC $\sim$2 Gyr old star clusters (blue crosses; \protect\citealt{Mucciarelli2008}). No corrections for non-local thermodynamic equilibrium effects were included for these elements. The black dashed line shows the solar abundance level.}
    \label{fig:Ti_com}
\end{figure*}

Comparing our abundances for the NGC 1846 sample to past studies of the LMC field stars (bar and inner disc) \citep{VanderSwaelmenM.HillV.2013} and the LMC intermediate-age cluster stars \citep{Mucciarelli2008} in Figure \ref{fig:Na_com} and \ref{fig:Ti_com}, our results indicate that the chemical composition of NGC~1846 is consistent with that of the LMC, resembling more the LMC disc than the bar. This is in line with what we know about NGC 1846, since this cluster is kinematically an LMC disk object \footnote{The line-of-sight velocity of NGC 1846 ($v_{\rm rad}$ $\sim$240 \kms; \citealt{MacKey2013}) is comparable with the LMC field's line-of-sight velocity at the position angle of the cluster with the prediction of disk rotation \citep{vanderMarel2002}.}. This also indicates that all the NGC 1846 stars observed are 1G rather than 2G.

Our results also agree with abundances from other LMC intermediate-age clusters (NGC 1651, 1783, 1978 and 2173), with the exception of V and O.  However, our NGC~1846 O abundance measurements interestingly show that they are located in the same region as the bar sample rather than with the disk. This could be due to the NLTE corrections that we used for our O abundance analysis that \cite{VanderSwaelmenM.HillV.2013} did not. In addition, they were only able to measure [O/Fe] in a very small number of disk stars, thus it is unclear if our abundance difference is significant or not.

\section{Conclusions}

We present detailed C, O, Na, Mg, Si, Ca, Ti, V, Zr, Ba, and Eu abundance measurements for 20 RGB stars in the LMC star cluster NGC 1846. This cluster is 1.95 Gyr old and lies just below the supposed lower age limit (2 Gyr) for the presence of multiple populations in clusters. Our measurements are based on high and low-resolution VLT/FLAMES spectra combined with photometric data from HST. Corrections for non-local thermodynamic equilibrium effects are also included for O, Na, Mg, Si, Ca, Fe and Ba. Our results show that there is no significant evidence for multiple populations in this cluster based on the lack of star-to-star spread in the Na and O abundances. 
However, we do detect a significant carbon abundance spread, indicating varying amounts of evolutionary mixing occurring on the RGB. This could be attributed to varying amounts of rotation which alters the amount of mixing from star to star. The general abundance patterns for NGC 1846 are similar to those seen in previous studies of LMC clusters and field stars.

Past studies on multiple populations in intermediate age clusters with similar masses ($\sim10^{5} \Msol$)  have found differing results. \cite{Mucciarelli2008} showed using VLT/UVES spectra that 4 of these clusters (NGC 1651, 1783, 1978 and 2173) spanning 1.5 to 2 Gyr in age (\citealt{Goudfrooij2014}; \citealt{Martocchia2018}) do not display obvious chemical inhomogeneities. However, in the recent decade, at least two studies have shown that some of these clusters display evidence for MPs. One example is found in the \cite{Saracino2020} study, where Na variations ($\sim 0.07$\,dex) have been found using VLT/MUSE spectra in NGC 1978, which has a similar age to NGC 1846. Furthermore, \cite{Kapse2022} recently found that NGC 2173 ($\sim 1.7$ Gyr) also exhibits light-element abundance variations using HST photometry. Therefore, in light of our findings, further high-res spectroscopic analysis is required to confirm the above-mentioned results. Hence, the jury is still out on the extent to which detections of putative multiple populations in younger systems match the variations seen in ancient clusters. 

\section*{Acknowledgements}

This paper includes data gathered with the 8m VLT
located at Cerro Paranal, Chile, and is based on observations collected at the European Southern Observatory under ESO programme 082.D-0387. 
ADM acknowledges Australian Research Council grant FT160100206.
This research was supported by the Australian Research Council Centre of Excellence for All Sky Astrophysics in 3 Dimensions (ASTRO 3D), through project number CE170100013. 

\section*{Data Availability}

The data used in this study are available in the ESO archive (https://archive.eso.org/eso/eso\_archive\_main.html) under program ID 082.D-0387. Our coadded spectra are available upon request. The ACS/WFC photometric data of NGC 1846 targets only were obtained from \cite{MacKey2013}; they originate in HST program GO9891 (PI: Gilmore) and GO10595 (PI: Goudfrooij).



\bibliographystyle{mnras}
\bibliography{NGC1846} 




\appendix

\section{Additional figures}
\begin{figure*}
	\includegraphics[width=\columnwidth]{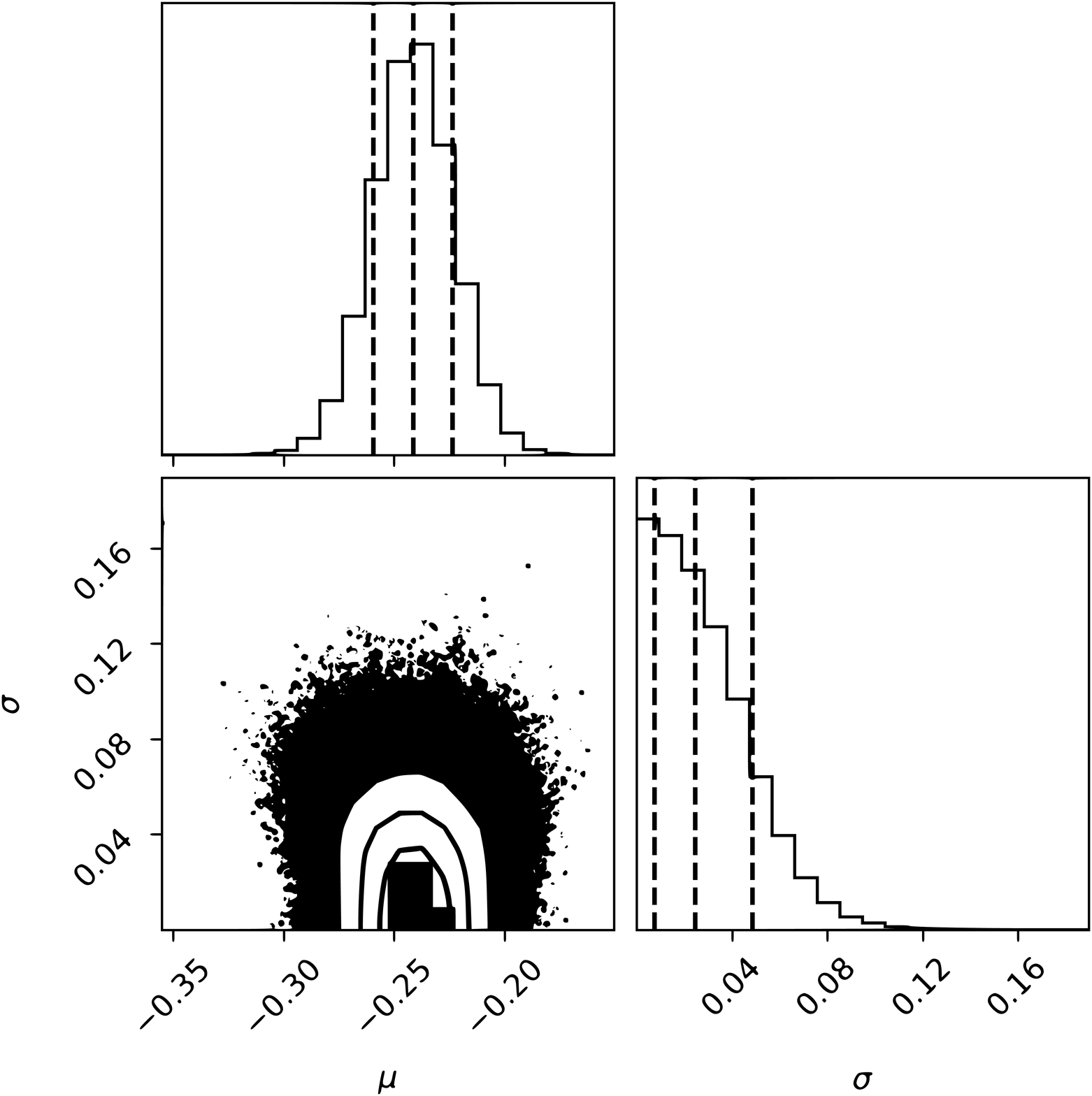}
	\includegraphics[width=\columnwidth]{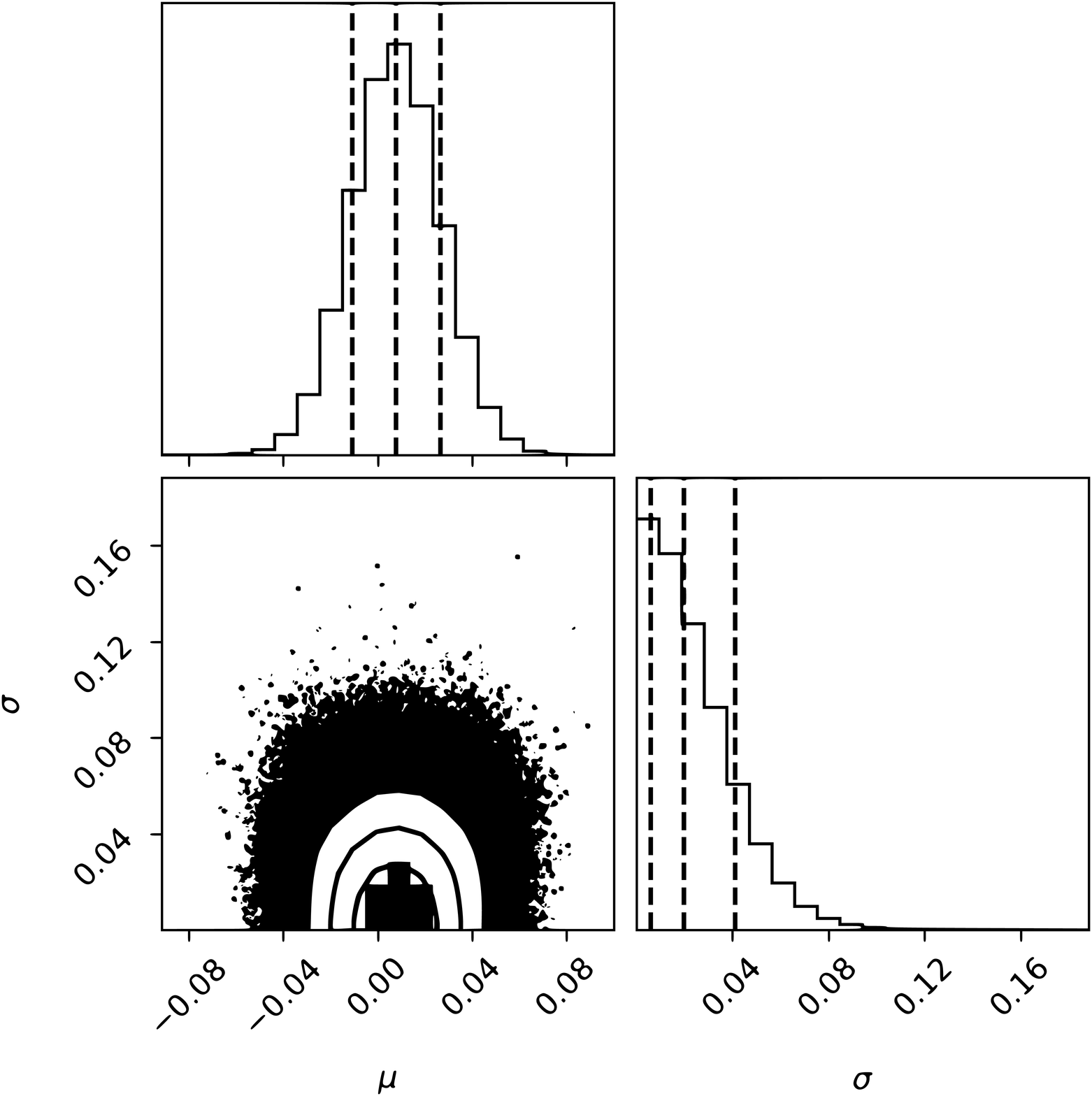}
	\includegraphics[width=\columnwidth]{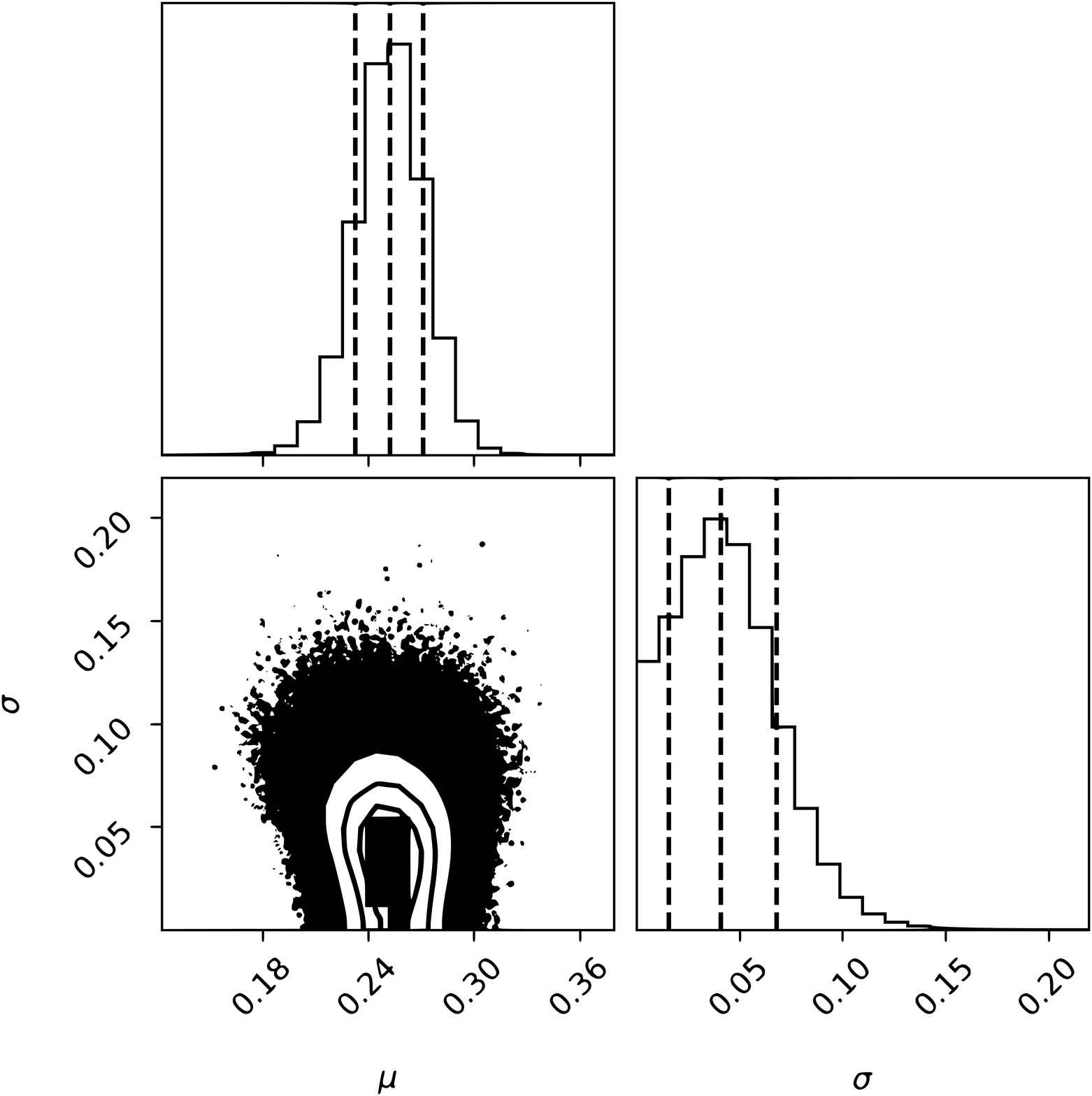}
    \caption{Corner plots showing the mean abundance and abundance dispersion of Na (top-left), Mg (top-right) and O (bottom) in NGC 1846. The dashed lines indicate the 25, 50 and 75 percentiles of the intrinsic spread of the different elements.}
    \label{fig:corners}
\end{figure*}

\section{Additional tables}
We show in Table \ref{tab:slope} the slopes of $\Teff$-[X/Fe] for our raw abundance measurements. We applied calibrations to all the raw abundance measurements except for carbon.

In Table \ref{tab:complete}, we show calibrated abundances of the measured elements. In the online version, we also provide another table with the raw abundances in the same format.
\begin{table}
	\centering
	\caption{Slopes of $\Teff$-[X/Fe] for raw abundance measurements for each element. }
	\label{tab:slope}
	\begin{tabular}{lr} 
		\hline
		Element & Slope (dex/1000\,K)\\
		\hline
		$\rm [C/Fe]$    & 0.454\\
		$\rm [O/Fe]$	& $-0.097$\\
		$\rm [Na/Fe]$	& 0.027\\
		$\rm [Mg/Fe]$	& $-0.022$\\
		$\rm [Si/Fe]$	& 0.126\\
		$\rm [Ca/Fe]$	& $-0.070$\\
		$\rm [Ti/Fe]$	& 0.022\\
		$\rm [Fe/H]$    & 0.146\\
		$\rm [V/Fe]$    & 0.010\\
		$\rm [Zr/Fe]$	& $-0.170$\\
		$\rm [Ba/Fe]$	& $-0.136$\\
		$\rm [Eu/Fe]$	& $-0.290$\\
		\hline
	\end{tabular}
\end{table}

\begin{landscape}
\begin{table}
	\centering
	\caption{Abundance table showing the calibrated abundance measurement for each star. The random and systematic errors are given for each element. }
	\label{tab:complete}
	\begin{tabular}{lcccccccccccc} 
		\hline 
		Name &
		$\rm[Fe/H]_\text{calib}$ &
		$\rm[C/Fe]_\text{calib}$ &
	     $\rm[O/Fe]_\text{calib}$ & $\rm[Na/Fe]_\text{calib}$ &  $\rm[Mg/Fe]_\text{calib}$ &  $\rm[Si/Fe]_\text{calib}$ \\ & $\rm[Ca/Fe]_\text{calib}$ &  $\rm[Ti/Fe]_\text{calib}$ & $\rm[V/Fe]_\text{calib}$ &  $\rm[Zr/Fe]_\text{calib}$ &  $\rm[Ba/Fe]_\text{calib}$ & $\rm[Eu/Fe]_\text{calib}$\\
		\hline
ACS-001 & 
$-0.565 \pm0.008 \pm0.100$ & 
$0.094 \pm0.002 \pm0.139$ & 
$0.112 \pm0.032 \pm0.068$ & 
$-0.299 \pm0.031 \pm0.075$ & 
$-0.053 \pm0.069 \pm0.078$ & 
$-0.063 \pm0.100 \pm0.085$ \\& 
$-0.059 \pm0.019 \pm0.077$ & 
$0.106 \pm0.023 \pm0.020$ & 
$-0.346 \pm0.008 \pm0.062$ & 
$0.040 \pm0.025 \pm0.062$ & 
$0.400 \pm0.007 \pm0.051$ & 
$0.246 \pm0.050 \pm0.062$\\

ACS-013 & 
 $-0.602 \pm0.008 \pm0.087$ & $0.325 \pm0.004 \pm0.033$ & $0.247 \pm0.020 \pm0.057$ & $-0.122 \pm0.017 \pm0.066$  & $0.071	\pm0.067 \pm0.070$ & $0.277 \pm0.053 \pm0.067$ \\& $-0.218	\pm0.019 \pm0.029$ & $-0.121 \pm0.024 \pm0.037$ & $-0.372 \pm0.005 \pm0.094$ &  $-0.234	\pm0.016 \pm0.092$	& $0.349 \pm0.001 \pm0.083$ & $0.325 \pm0.064 \pm0.055$\\

ACS-017 & 
$-0.554 \pm0.012 \pm0.086$ & $0.091 \pm0.007 \pm0.028$ & $0.223 \pm0.022 \pm0.057$ & $-0.223 \pm0.017 \pm0.066$  & $-0.027	\pm0.040 \pm0.051$ & $0.102 \pm0.049 \pm0.064$ \\& $-0.063	\pm0.017 \pm0.046$ & $0.039 \pm0.023 \pm0.036$ & $-0.132 \pm0.013 \pm0.094$ &  $0.047	\pm0.017 \pm0.081$	& $0.380 \pm0.011 \pm0.082$ & $0.312 \pm0.065 \pm0.050$\\

ACS-025 & 
$-0.580 \pm0.009 \pm0.077$ & $-0.140 \pm0.006 \pm0.013$ & $0.292 \pm0.031 \pm0.058$ & $-0.198 \pm0.017 \pm0.065$  & $0.035	\pm0.061 \pm0.055$ & $0.002 \pm0.067 \pm0.073$ \\& $-0.098	\pm0.020 \pm0.040$ & $0.062 \pm0.028 \pm0.028$ & $-0.228 \pm0.006 \pm0.103$ &  $0.000	\pm0.018 \pm0.093$	& $0.314 \pm0.013 \pm0.089$ & $0.397 \pm0.062 \pm0.052$\\																		

ACS-030 & 
$-0.631 \pm0.011 \pm0.079$ & $0.172 \pm0.006 \pm0.031$ & $0.398 \pm0.026 \pm0.047$ & $-0.196 \pm0.024 \pm0.00$  & $0.037	\pm0.050 \pm0.057$ & $0.079 \pm0.061 \pm0.070$ \\& $-0.147	\pm0.023 \pm0.040$ & $-0.111 \pm0.036 \pm0.037$ & $-0.367 \pm0.007 \pm0.090$ & $-0.191	\pm0.030 \pm0.092$	& $0.372 \pm0.015 \pm0.088$ & $0.445 \pm0.078 \pm0.048$\\

ACS-036 & 
$-0.555 \pm0.012 \pm0.078$ & 
$0.140 \pm0.013 \pm0.044$ & 
$0.252 \pm0.041 \pm0.050$ & 
$-0.263 \pm0.024 \pm0.066$  & 
$-0.057 \pm0.041 \pm0.049$ & 
$0.043 \pm0.054 \pm0.060$ \\& 
$-0.097 \pm0.021 \pm0.037$ & 
$-0.044 \pm0.035 \pm0.038$ & 
$-0.285 \pm0.007 \pm0.091$ & 
$-0.076 \pm0.026 \pm0.093$ & 
$0.281 \pm0.023 \pm0.106$ & 
$0.409 \pm0.085 \pm0.048$\\

ACS-043 & 
$-0.536 \pm0.010 \pm0.086$ & 
$0.228 \pm0.005 \pm0.030$ & 
$0.321 \pm0.034 \pm0.053$ & 
$-0.328 \pm0.027 \pm0.073$  & 
$-0.011 \pm0.050 \pm0.050$ & 
$0.085 \pm0.058 \pm0.064$ \\& 
$-0.106 \pm0.025 \pm0.036$ & 
$0.003 \pm0.035 \pm0.032$ & 
$-0.244 \pm0.008 \pm0.089$ & 
$-0.149 \pm0.037 \pm0.094$ & 
$0.294 \pm0.022 \pm0.101$ & 
$0.359 \pm0.108 \pm0.052$\\

ACS-046 & 
$-0.613 \pm0.012 \pm0.087$ & 
$-0.055 \pm0.006 \pm0.038$ & 
$0.048 \pm0.069 \pm0.070$ & 
$-0.249 \pm0.031 \pm0.074$  & 
$0.061 \pm0.057 \pm0.052$ & 
$0.128 \pm0.083 \pm0.071$ \\& 
$0.075 \pm0.031 \pm0.028$ & 
$0.042 \pm0.041 \pm0.035$ & 
$-0.394 \pm0.011 \pm0.097$ & 
$-0.246 \pm0.081 \pm0.096$ & 
$0.316 \pm0.020 \pm0.069$ & 
$0.164 \pm0.200 \pm0.059$\\

ACS-047 & 
$-0.541 \pm0.015 \pm0.082$ & 
$0.050 \pm0.003 \pm0.077$ & 
$0.309 \pm0.041 \pm0.055$ & 
$-0.203 \pm0.038 \pm0.067$  & 
$0.004 \pm0.055 \pm0.047$ & 
$0.011 \pm0.076 \pm0.067$ \\&
$-0.160 \pm0.036 \pm0.041$ & 
$-0.053 \pm0.057 \pm0.043$ & 
$-0.264 \pm0.010 \pm0.093$ & 
$0.032 \pm0.050 \pm0.090$ & 
$0.422 \pm0.024 \pm0.068$ & 
$0.333 \pm0.135 \pm0.049$\\

ACS-053  & 
$-0.676 \pm0.014 \pm0.076$ & 
$0.221 \pm0.006 \pm0.035$ & 
$0.286 \pm0.046 \pm0.047$ & 
$-0.673 \pm0.065 \pm0.075$  & 
$-0.122 \pm0.068 \pm0.047$ & 
$-0.004 \pm0.080 \pm0.066$ \\&
$-0.117 \pm0.035 \pm0.04$ & 
$-0.053 \pm0.047 \pm0.046$ & 
$-0.336 \pm0.012 \pm0.091$ & 
$-0.308 \pm0.096 \pm0.096$ & 
$0.221 \pm0.046 \pm0.118$ & 
$0.337 \pm0.194 \pm0.041$\\

ACS-059  & 
$-0.503 \pm0.021 \pm0.081$ & 
$0.202 \pm0.007 \pm0.028$ & 
$-0.122 \pm0.109 \pm0.064$ & 
$-0.296 \pm0.059 \pm0.078$  & 
$-0.014 \pm0.060 \pm0.050$ & 
$0.073 \pm0.097 \pm0.068$ \\&
$-0.143 \pm0.051 \pm0.044$ & 
$-0.029 \pm0.077 \pm0.047$ & 
$-0.228 \pm0.013 \pm0.099$ & 
$-0.314 \pm0.117 \pm0.107$ & 
$0.322 \pm0.037 \pm0.064$ & 
$-1.460 \pm5.316 \pm0.984$\\

ACS-066  & 
$-0.497 \pm0.019 \pm0.081$ & 
$0.115 \pm0.038 \pm0.042$ & 
$0.490 \pm0.036 \pm0.045$ & 
$-0.224 \pm0.053 \pm0.074$  & 
$0.011 \pm0.089 \pm0.054$ & 
$-0.206 \pm0.089 \pm0.076$ \\&
$-0.176 \pm0.047 \pm0.035$ & 
$-0.381 \pm0.083 \pm0.070$ & 
$-0.251 \pm0.014 \pm0.104$ & 
$0.047 \pm0.063 \pm0.103$ & 
$0.397 \pm0.031 \pm0.076$ & 
$0.299 \pm0.237 \pm0.052$\\

ACS-080  & 
$-0.624 \pm0.007 \pm0.091$ & 
$0.014 \pm0.006 \pm0.036$ & 
$0.272 \pm0.029 \pm0.063$ & 
$-0.332 \pm0.023 \pm0.079$  & 
$0.008 \pm0.046 \pm0.062$ & 
$0.023 \pm0.053 \pm0.076$ \\&
$-0.125 \pm0.017 \pm0.036$ & 
$0.024 \pm0.022 \pm0.027$ & 
$-0.327 \pm0.005 \pm0.096$ & 
$-0.100 \pm0.019 \pm0.087$ & 
$0.447 \pm0.012 \pm0.069$ & 
$0.469 \pm0.051 \pm0.057$\\

ACS-081  & 
$-0.599 \pm0.008 \pm0.086$ & 
$-0.153 \pm0.009 \pm0.038$ & 
$0.274 \pm0.025 \pm0.060$ & 
$-0.221 \pm0.015 \pm0.066$  & 
$0.010 \pm0.046 \pm0.057$ & 
$0.038 \pm0.060 \pm0.069$ \\&
$-0.149 \pm0.018 \pm0.039$ & 
$-0.038 \pm0.025 \pm0.029$ & 
$-0.307 \pm0.005 \pm0.090$ & 
$-0.119 \pm0.018 \pm0.085$ & 
$0.383 \pm0.013 \pm0.083$ & 
$0.503 \pm0.056 \pm0.052$\\

ACS-082  & 
$-0.674 \pm0.009 \pm0.08$ & 
$0.275 \pm0.007 \pm0.019$ & 
$0.254 \pm0.019 \pm0.058$ & 
$-0.214 \pm0.018 \pm0.065$  & 
$0.059 \pm0.042 \pm0.052$ & 
$0.242 \pm0.067 \pm0.059$ \\&
$-0.121 \pm0.022 \pm0.044$ & 
$-0.068 \pm0.028 \pm0.032$ & 
$-0.375 \pm0.006 \pm0.094$ & 
$-0.121 \pm0.024 \pm0.089$ & 
$0.283 \pm0.021 \pm0.102$ & 
$0.570 \pm0.057 \pm0.047$\\

ACS-085  & 
$-0.610 \pm0.009 \pm0.083$ & 
$0.241 \pm0.007 \pm0.030$ & 
$0.306 \pm0.038 \pm0.059$ & 
$-0.359 \pm0.021 \pm0.076$  & 
$0.121 \pm0.088 \pm0.058$ & 
$0.251 \pm0.049 \pm0.057$ \\&
$-0.112 \pm0.022 \pm0.036$ & 
$-0.058 \pm0.029 \pm0.033$ & 
$-0.320 \pm0.006 \pm0.099$ & 
$-0.186 \pm0.023 \pm0.098$ & 
$0.319 \pm0.017 \pm0.090$ & 
$0.340 \pm0.083 \pm0.055$\\

ACS-090  & 
$-0.595 \pm0.011 \pm0.082$ & 
$-0.071 \pm0.005 \pm0.027$ & 
$0.264 \pm0.044 \pm0.053$ & 
$-0.204 \pm0.030 \pm0.063$  & 
$-0.063 \pm0.042 \pm0.054$ & 
$0.167 \pm0.065 \pm0.065$ \\&
$-0.094 \pm0.025 \pm0.045$ & 
$-0.083 \pm0.032 \pm0.037$ & 
$-0.310 \pm0.007 \pm0.084$ & 
$-0.113 \pm0.031 \pm0.086$ & 
$0.312 \pm0.016 \pm0.084$ & 
$0.457 \pm0.072 \pm0.046$\\

ACS-092  & 
$-0.575 \pm0.011 \pm0.083$ & 
$0.137 \pm0.007 \pm0.032$ & 
$0.234 \pm0.030 \pm0.051$ & 
$-0.265 \pm0.020 \pm0.073$  & 
$-0.096 \pm0.104 \pm0.053$ & 
$0.176 \pm0.078 \pm0.064$ \\&
$-0.147 \pm0.025 \pm0.035$ & 
$-0.053 \pm0.040 \pm0.034$ & 
$-0.409 \pm0.009 \pm0.098$ & 
$-0.350 \pm0.040 \pm0.104$ & 
$0.389 \pm0.018 \pm0.071$ & 
$0.254 \pm0.112 \pm0.053$\\

ACS-102  & 
$-0.610 \pm0.015 \pm0.079$ & 
$0.091 \pm0.007 \pm0.017$ & 
$0.118 \pm0.076 \pm0.054$ & 
$-0.142 \pm0.028 \pm0.067$  & 
$0.091 \pm0.058 \pm0.046$ & 
$0.271 \pm0.104 \pm0.055$ \\&
$-0.113 \pm0.034 \pm0.036$ & 
$-0.075 \pm0.061 \pm0.046$ & 
$-0.473 \pm0.014 \pm0.099$ & 
$-0.213 \pm0.073 \pm0.098$ & 
$0.405 \pm0.031 \pm0.072$ & 
$0.416 \pm0.114 \pm0.046$\\

ACS-112  & 
$-0.651 \pm0.014 \pm0.082$ & 
$0.037 \pm0.005 \pm0.050$ & 
$0.092 \pm0.052 \pm0.051$ & 
$-0.380 \pm0.030 \pm0.075$  & 
$0.173 \pm0.096 \pm0.056$ & 
$0.304 \pm0.095 \pm0.052$ \\&
$0.011 \pm0.038 \pm0.049$ & 
$0.143 \pm0.051 \pm0.027$ & 
$-0.390 \pm0.012 \pm0.095$ & 
$-0.139 \pm0.067 \pm0.089$ & 
$0.376 \pm0.033 \pm0.061$ & 
$0.248 \pm0.200 \pm0.049$\\

		\hline
	\end{tabular}
	\end{table}
\end{landscape}


\bsp	
\label{lastpage}
\end{document}